\documentclass[12pt,3p]{llncs}

\usepackage[sectionbib]{natbib}

\usepackage{bm}
\usepackage{comment}
\usepackage{amsmath}
\usepackage{amsbsy}
\usepackage{amssymb}
\usepackage{amscd}
\usepackage{amsfonts}
\usepackage{mathrsfs,amsmath} 
\usepackage{graphicx} 
\usepackage{graphics}
\usepackage{epstopdf}
\usepackage{enumitem}
\usepackage{xcolor}
\usepackage{subfigure}
\usepackage{setspace}
\usepackage{pifont}
\usepackage{textalpha}

\usepackage{hyperref}

\usepackage{subfigure}
\usepackage{float}
\usepackage{times}
\usepackage{fancyhdr}
\usepackage{fancyheadings}
\usepackage{rotating}
\usepackage{pstricks}
\usepackage{soul}
\definecolor{Gray}{gray}{0.85}
\usepackage[top=30mm, bottom=25mm, left=20mm, right=20mm]{geometry}
\allowdisplaybreaks

\definecolor{Lightgray}{RGB}{235,235,235}
\definecolor{Gray2}{gray}{0.9}
\definecolor{Gray}{gray}{0.85}
\definecolor{Gray3}{gray}{0.8}
\usepackage{enumitem}
\usepackage{amsbsy}
\usepackage{soul}
\usepackage{float}
\usepackage{enumitem}
\usepackage{mathrsfs}
\usepackage{mdwlist}
\usepackage{color}
\usepackage{soul}
\usepackage{url}

\usepackage{caption}

\usepackage{longtable}

\usepackage{multirow}
\usepackage{booktabs}
\usepackage{array}
\usepackage{tabu}
\usepackage{mathtools}
\usepackage{cancel}

\newcommand{\otoprule}{\midrule[\heavyrulewidth]} 

\usepackage[normalem]{ulem}               
\newcommand\redout{\bgroup\markoverwith
	{\textcolor{red}{\rule[0.5ex]{2pt}{0.8pt}}}\ULon}

\renewcommand{\figurename}{\footnotesize{Fig.}}

\usepackage{fancyhdr}            
\fancypagestyle{plain}{%
	\fancyhead{}                                     
	\fancyfoot[CE,CO]{\thepage}       
	\fancyhead[CE,CO]{\textcolor[rgb]{0.64,0.15,0.15}{Published in \emph{Extreme Mechanics Letters} (2020),  \textbf{134}, 103741 
			\\
			doi: \href{https://doi.org/10.1016/j.eml.2020.100977 }{10.1016/j.eml.2020.100977 }
	}}
	\setlength{\headheight}{14.5pt}}

\begin{document}

\title{Design of tunable acoustic metamaterials \\ with periodic piezoelectric  microstructure }

\author{ Andrea Bacigalupo\inst{1}, Maria Laura De Bellis\inst{2}, Diego Misseroni\inst{3}}
\institute{ University of Genoa, Department DICCA, via Montallegro 1, Genoa, Italy  \and  University of Chieti-Pescara, Department INGEO, Viale Pindaro 42, Pescara, Italy \and University of Trento, via Mesiano 77, Trento, Italy }

\maketitle

\begin{abstract}
\textcolor{black}{An innovative special class of tunable periodic metamaterials is designed, suitable for realising high-performance  acoustic filters.} The metamaterial is made up of a phononic crystal coupled to local resonators. Such local resonators consist of masses enclosed into piezoelectric rings, shunted by either dissipative or non-dissipative electrical circuit.
By tuning the impedance/admittance of such electrical circuits, it is possible to fully adjust the constitutive properties of the shunting piezoelectric material. 
This feature paves the way for unconventional behaviours, well beyond the capabilities achievable with classical materials. It follows that the acoustic properties of the periodic metamaterial can be adaptively modified, in turn, opening new possibilities for the control of pass and stop bands. By exploiting a generalization of the Floquet-Bloch theory, the in-plane free wave propagation in the tunable metamaterial is investigated, by varying a certain tuning parameter, to show the efficiency of the proposed shunting piezoelectric system as a wave propagation control device.
Particular attention is devoted to the determination of the in-plane constitutive equations of the shunting piezoelectric phase in the transformed Laplace space. 
Finally,
broad design directions of tunable acoustic filters aiming to a changing performance requirement in real-time, is also provided.
\end{abstract}

\begin{keywords}
 periodic microstructure, piezoelectric shunting,
  tunable metamaterials, 
  Bloch wave propagation, band gap control
\end{keywords}

\section{Introduction}

The study of metamaterials is increasingly emerging as a cutting edge interdisciplinary area, including physics, material science and engineering.
Metamaterials are engineered composites, specifically tailored to exhibit outstanding constitutive properties,  
well above those achievable with classical materials.
First introduced in optics
and photonics \citep{Pendry2006,PhysRevLett.84.4184}, metamaterials have afterwards established themselves in the fields of elastodynamics and acoustics \citep{ziolkowski2001wave,PhysRevB.71.014103,Fang2006} for a wide range of intriguing applications ranging from filtering, to wave-guiding, self-collimation, mechanical energy transfer, wave polarization up to band-gaps control, i.e. more generally manipulation of the dispersive properties of vibrational waves \citep{PhysRev.134.A158,
Langley1996,ostoja2002lattice,Movchan2003,
Ruzzene_2003,Tee2010,dcemmie2011waves,allegri2013wave,
lemoult2013wave,de2017auxetic,
Bacigalupo2017a,bacigalupo2017wave,PICCOLROAZ2017152,
BACIGALUPO2018183,
DAlessandro2018,bacigalupo2019complex,2Bordiga2019,Dalcorso2019,
KAMOTSKI2019292,Park2019StudyOA,guo2019frequency,bacigalupo2020chiral}. 
\textcolor{black}{
By focusing on this latter context, acoustic metamaterials have being designed from periodic distributions of inclusions (or scatterers) embedded in a matrix, i.e. phononic crystals, with the addition of local resonators which enable unique sub-wavelength properties to emerge accordingly with \cite{LU200934,liu2020review}.} The core idea behind their design is to engineer
the architecture and the geometry of their microstructure at different scales of interest in order to achieve unusual macroscopic properties. 
In order to 
meet the requirement of optimal design,  parametric and topological optimization techniques can be successfully used \citep{diaz2005design,bacigalupo2016optimal,
ranjbar2016vibroacoustic,WANG2017250,
bacigalupo2019machine,bruggi2019optimal,choi2019optimal,
kumar2019isogeometric,RONG2019819}.
 Within the framework of acoustic metamaterials, some of the limitations of standard materials such as negative refraction \citep{Liu2011b,Christensen2012,Morini2019,
 2Bordiga2019,Bordiga2019}, superlenses \citep{Yan2013,Park2015,Brun2019a}, and invisibility of defects embedded into both lattice and continuous systems \citep{Brun2009,Norris2011,Colquitt2014,Misseroni2016,
stenger2012}  can be overcome.   
These astonishing features, unachievable with natural materials, have found application in the fabrication of new devices such as concentration detectors, vibration dampers and also in the protection of buildings from earthquakes \citep{Brun2012,brun2013phononic,
Brule2014,ACHAOUI201630,Achaoui_2017,CARTA2016216,colombi2016seismic,Miniaci2016,
craster2018elastic,ungureanu2019influence}.
Different techniques have been proposed to accomplish the aforementioned targets.  Some of these include the change of the mechanical properties of the material, such as mass density, inertia and stiffness \citep{parnell2011nonlinear,
Colquitt2014,kadic2014pentamode,zigoneanu2014three,
darabi2018experimental,Misseroni2019}, others 
the insertion of spatially-local inertial or Helmholtz resonators within either lattice materials or microstructured continuum materials \citep{Liu2011a,Zhou2012,Bigoni2013,Bacigalupo2016a,
Cummer2016,Mae1501595,
Zhou2016,kaina2017slow,LEPIDI2018186}. \\
More sophisticated techniques are based on the use of active phases exploiting multi-field couplings, such as the electro- or the magneto-mechanical one, to achieve the  wave propagation control. These approaches leave wide freedom of design without changing the mass of the system. In particular, the pioneering work of  \cite{Forward1979} demonstrated the efficiency of piezoelectric elements connected to electrical circuits for vibration control. More specifically, one way to couple the mechanical and the electrical fields concerns the use of piezoelectric patches shunted by 
electrical networks (\textit{shunted piezoelectric phases}) in both active and passive control schemes. 
Contrary to active control, in passive schemes, a sensing element is not necessary.  \textcolor{black}{In most cases,
the metastructure consists of an array of passive electrical circuits mounted on the vibrating structure, each of them directly connected to the piezoelectric device \citep{HAGOOD1991,Hollkamp1994,preumont1997vibration,
Thomas2009,Wang_2011,bergamini2014phononic,ZHANG2015104,Li_2018}}. 
A detailed review of different shunt piezoelectric systems for noise, vibration and wave propagation control is provided in \cite{chen2018review,Marakakis2019,zangeneh2019active}.                                                   
\textcolor{black}{The case of metamaterials with piezoelectric phases shunted by dissipative and/or non-dissipative electrical circuit, characterized by variable capacitors, is 
widely investigated in the spectral design of waveguides and/or acoustic filters \citep{
thorp2001attenuation,Airoldi_2011,
casadei2012piezoelectric,collet2012structural,beck2013power,beck2014response,
chen2016adaptive,
ouisse2016piezo,Zhu2016,Xu2018,
li2019active}.}\\
Within this context, in the present paper the constitutive properties 
of a tunable metamaterial, made of a piezoelectric phase coupled to a generic electrical circuit with a certain equivalent admittance, are consistently determined  both in the physical space and in the frequency transformed space.
 More specifically, in the case of generic dissipative electrical circuits the constitutive relations are in general frequency-dependent. \textcolor{black}{On the other hand, with regard to non-dissipative electrical circuits,
when they are characterized by i) purely capacitive equivalent admittance, frequency-independent constitutive relations are obtained; ii) purely inductive equivalent admittance, frequency-dependent constitutive relations are obtained.}\\
In particular, with the aim of designing high-performance tunable acoustic filters for the passive wave propagation control, a specific class of periodic metamaterials with a phase shunted by either a dissipative or non dissipative electrical circuit is here proposed. The three phase metamaterial is characterized by a phononic crystal coupled to local resonators. The phononic crystal is made by an in-plane periodic tessellation of rigid and heavy external rings enclosed within a soft and light matrix, while the local resonators consist of rigid and heavy internal disks connected to the external rings through inner rings  made of a piezoelectric material shunted by an electrical circuit.
 The main idea is to properly modifying the equivalent impedance/admittance of the electrical circuit in order to tune the constitutive properties of the shunting piezoelectric phase. This means that, within the design of the metamaterial, its overall constitutive properties can be fully tailored by adjusting parameters of the electrical circuit, without modifications of the microstructural geometry and/or of the non shunted material properties.
As a consequence, a control of the band structure of the periodic metamaterial is performed, in terms of amplitude and central frequency of pass and stop bands. More specifically, focus is on a non dissipative electrical circuit, i.e. an adaptive capacitor with both positive and negative capacitance, leading to a real-valued Floquet-Bloch spectrum. A critical investigation of the metamaterial behaviour,  
as a a set of design geometric parameters change,  is useful to provide broad guidelines to achieve optimal design solutions.\\
\begin{figure}[h]
\centering
\includegraphics[width=1\textwidth]{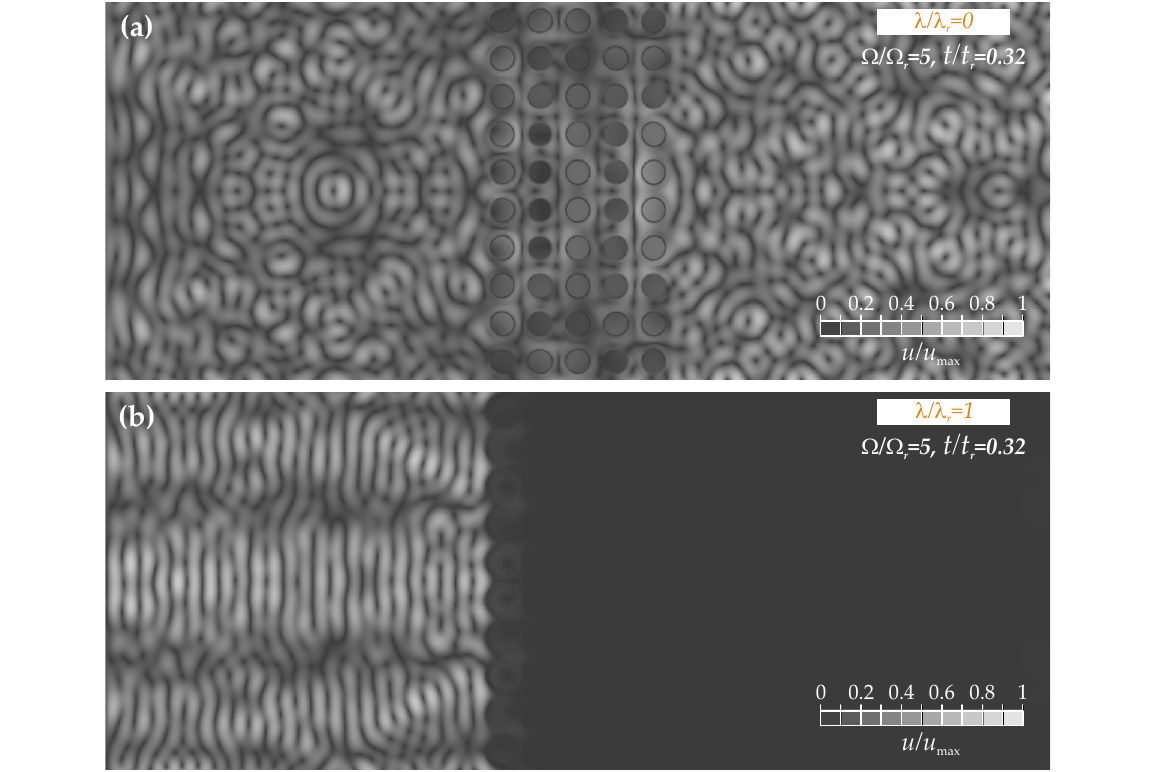}
\caption{\footnotesize  \textcolor{black}{Filtering performance of the tunable metamaterials. A rectangular strip of homogeneous material, with a central core made by a portion of the metamaterial, undergoes a mono-frequency time-harmonic displacement excitation.  (a) case of excitation frequency falling into a pass band; (b) case of excitation frequency falling into
a stop band.}}
\label{figure0}
\end{figure}
\textcolor{black}{In order to test the effectiveness of the periodic tunable metamaterial
as acoustic filter, a numerical experiment has been performed on a thin rectangular strip of homogeneous material in which a central strip has been replaced by a portion of the metamaterial reported in Fig.\ref{figure0}. 
A mono-frequency time-harmonic
displacement excitation is imposed on the left side of the specimen. By properly tuning the parameters characterising the electrical circuit, the frequency band-structure is modified. In the case the excitation frequency
falls into a pass band no filtering properties are exhibited as in Fig. 1(a), where it emerges that the wave
front passes through the microstructured core without significant reduction in the vibration amplitude.
On the other hand, when the excitation frequency falls into a stop band, the tunable metamaterial behaves
as a highly efficient acoustic filter as in Fig. 1(b), in which the wave front is nearly entirely reflected and
trapped in the microstructured core.}\\
The paper is organized as follows. In Section \ref{S2}
the periodic material with a piezoelectric phase shunted by an electrical circuit is described. In Section \ref{S2}.1
the in-plane constitutive equations of the orthotropic shunted piezoelectric phase are derived both for either an electrical circuit characterized by generic equivalent admittance or for one 
characterized by purely capacitive equivalent admittance. Section \ref{S2}.2
is devoted to the dynamic governing equations of the periodic material with a shunted piezoelectric phase.
 Within the framework of a first-order continuum,  the governing equation of the in-plane free Bloch wave propagation together with the Floquet-Bloch boundary conditions, imposed on the periodic cell, are defined.
\textcolor{black}{In Section \ref{Section31} numerical experiments are devoted to demonstrate the effectiveness of the pass and stop bands tunable control of the proposed shunted metamaterial.
Section \ref{Section41} focuses on the performance as a filter of the designed tunable metamaterial in the framework of a numerical experiment.}
Finally, some concluding remarks are summarized in Section \ref{S5}. \\

\section{Tunable metamaterials with  piezoelectric shunting} \label{S2}
\begin{figure}[hbtp]
\centering
\includegraphics[width=1\textwidth]{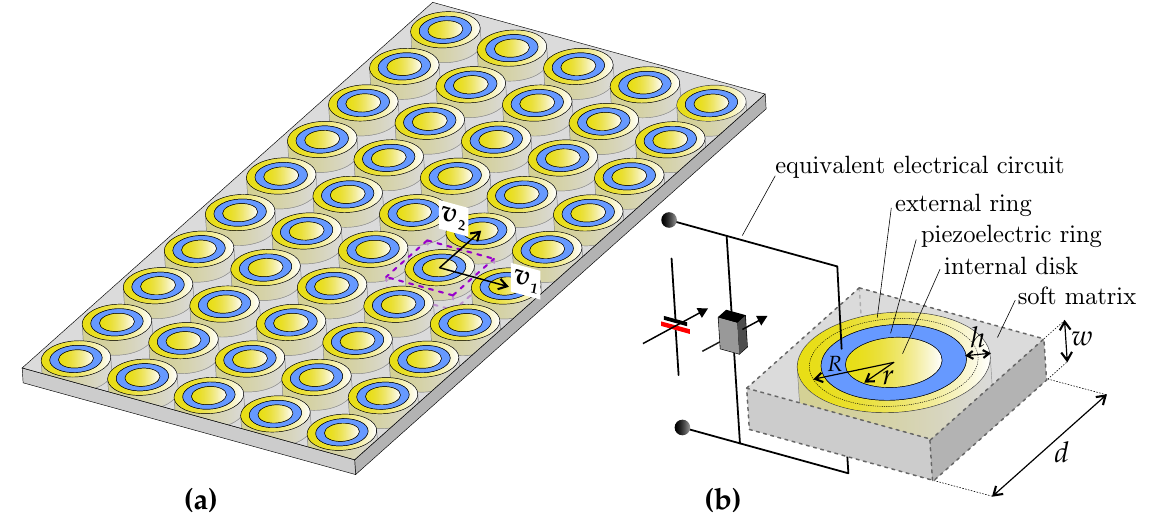}
\caption{\footnotesize \textcolor{black}{(a) A generic portion of the heterogeneous material; (b) Scheme of the Periodic Cell containing the shunted piezoelectric material.}}
\label{figurePrima}
\end{figure}

\noindent We focus on an infinite periodic three-phase microstructured metamaterial, a generic portion of which is schematically depicted in Fig. \ref{figurePrima}(a).
\noindent Such periodic material is characterized by 
the in-plane regular repetition of a periodic cell, see Fig. \ref{figurePrima}(b), along 
two orthogonal periodicity vectors $\textbf{v}_1 = d \textbf{e}_1$ ,
$\textbf{v}_2 = d \textbf{e}_2$, with $\textbf{e}_1$ and $\textbf{e}_2$ being a given orthogonal base.
\textcolor{black}{The periodic cell $\mathfrak{A}$, of thickness $w$, is made of an external soft and light matrix embedding both an outer rigid and heavy ring (with mean radius $R$)
	and an inner ring made of a shunted piezoelectric material of thickness $h$.}
The geometry is  completed by a circular disk, of radius $r$, made of the same  material as the outer ring.   
In this configuration each piezoelectric ring is connected in parallel through electrodes to an external electrical circuit with tunable equivalent impedance/admittance, see Fig. \ref{figurePrima}(b). In this framework, 
the constitutive properties of the piezoelectric shunting can be modified by varying such equivalent impedance/admittance  to control the spectral properties of the acoustic metamaterial. In this way the stop- and pass-band frequency can be properly shifted in order to filter Bloch waves propagating in the material at a given frequency.\\
In the following we will focus first on the treatment of the constitutive properties that characterize the piezoelectric shunting phase  (Section 2.1). Then the field equations governing the dynamic behaviour of the metamaterial will be introduced, consistently with the constitutive equations of both the  shunted piezoelectric phase and the other characterized by linear elastic behaviour (Section 2.2).

\begin{table}[H]
	\footnotesize
	\renewcommand{\tablename}{\footnotesize{Table}}
	\centering
	\caption*{\footnotesize{\textcolor{black}{\textbf{List of Symbols}}}}
	\label{Tab_Summary}
	\textcolor{black}{
	\begin{tabu}  to 0.95\linewidth {*{1}{X[0.25c]}*{1}{X[0.93l]}|*{1}{X[0.25c]}*{1}{X[0.93l]}}
		\otoprule  
		Symbol   &  \centering Explanation&  Symbol &\centering Explanation\\
		\midrule
		$\mathcal{L} $ & Bilateral Laplace transform operator & $\textbf{m}$& Outward normal unit vector \\   
		$\mathcal{L}^{-1} $ & Inverse bilateral Laplace transform operator& $\Gamma$& Closed polygonal curve \\ 
		$\mathcal{F}^{-1}$ & Inverse Fourier transform operator& $\Xi$& Curvilinear abscissa    \\
		$*$& Convolution product &$\Xi_j$& Vertices of $\Gamma$  \\ 
		$\textbf{e}_j$& j$^{th}$ unit vector of the canonical basis &$t$& Temporal variable  \\ 
		$\iota$& Imaginary unit &$s$& Complex angular frequency   \\ 
		$\delta$&  Dirac delta function &$\omega$& Angular frequency \\ 
		$\mathfrak{A}$&  Periodic cell &$R$, $r$, $h$,  $w$ & Geometric dimensions of the microstructure  \\ 
		$\partial \mathfrak{A}$&  Boundary of  $\mathfrak{A}$ &$u_j$& displacement field component  \\ 
		$\partial \mathfrak{A}^{\pm}$ &  Positive/negative part of $\partial \mathfrak{A}$ &$\varepsilon_{ij}$&  Strain tensor component \\ 
		$d$&  Edge of the periodic cell   &$\phi$& electric potential field \\ 
		$\textbf{v}_j$& j$^{th}$ periodicity vector &$\sigma_{ij}$& Stress tensor component\\ 
		$\textbf{x}$& Position vector  &$D_j$& electric displacement component \\
		$\textbf{x}^{\pm}$& Position vector of the point $P$ $\in \partial \mathfrak{A}^{\pm}$ &$V_3$& Electric potential difference in $\textbf{e}_3$ \\
		$x_j$&  component of $\textbf{x}$  &$I_3$ & Current in $\textbf{e}_3$ \\
		$\mathfrak{B}$& Dimensionless first Brillouin zone   & $C_{ijhk}$& Elasticity tensor component\\ 
		$\textbf{k}$& Wave vector  &$\beta_{ij}$& Dielectric permittivity tensor component \\ 
		$k_j$&  component of  $\textbf{k}$  &${e}_{ijk}, \, \widetilde{e}_{ijk}$ & Stress-charge coupling tensor component  \\ 
		$\rho$& Mass density&$\mathord{\buildrel{\lower3pt\hbox{$\scriptscriptstyle\frown$}} \over g}$&  Laplace transformed function of $g$\\
		$\mathord{\buildrel{\lower3pt\hbox{$\scriptscriptstyle\frown$}} \over g}^{\pm}$&  Value of $\mathord{\buildrel{\lower3pt\hbox{$\scriptscriptstyle\frown$}} \over g}$ in  $\textbf{x}^{\pm}$ & 
		$\mathord{\buildrel{\lower3pt\hbox{$\scriptscriptstyle\frown$}} \over b_j}$& component of the transformed source term \\ 
		$ A^{(P)}$& Area of the piezoelectric annulus &  
		$L^{(P)} $& Dimension of the piezoelectric phase in $\textbf{e}_3$\\ 
		$Y^{(EL)}_{33}$& Total equivalent admittance &  
		$Y^{(P)}_{33}$& Equivalent admittance of the piezoelectric materia   \\ 
		$Y^{(SU)}_{33}$& Equivalent shunting admittance &  
		$Z^{(EL)}_{33}$& Total equivalent impedance   \\ 
		$\lambda(s)$ & Tuning function &  
		$\lambda$ & Tuning parameter    \\ 
		$\lambda_R$ & Resonance value of the tuning parameter &  
		$\lambda^*$ & Shifted tuning parameter    \\ 
		$\mathcal{C}$ & Capacitance & 
		$C_{ijhk}^{EL}$&  Equivalent elastic tensor component of the shunted piezoelectric phase   \\ 
		$\beta_{ij}^{EL}$& Equivalent dielectric permittivity tensor component of the shunted piezoelectric phase&  
		$C_{ijhk}^{EL0}$& Value of $C_{ijhk}^{EL}$ for $\lambda$=0  \\ 
		$C_{ijhk}^{EL \infty}$& Value of $C_{ijhk}^{EL}$ for $\lambda\to \pm \infty$ &  
		$C^\textrm{\ding{71}}_{ijkl}$ & Equivalent elastic tensor component of the tunable metamaterial  \\ 
	\end{tabu}}
\end{table}

\newpage

\subsection{Shunted piezoelectric constitutive model}\label{Section3}

\begin{figure}[H]
   \includegraphics[width=1\textwidth]{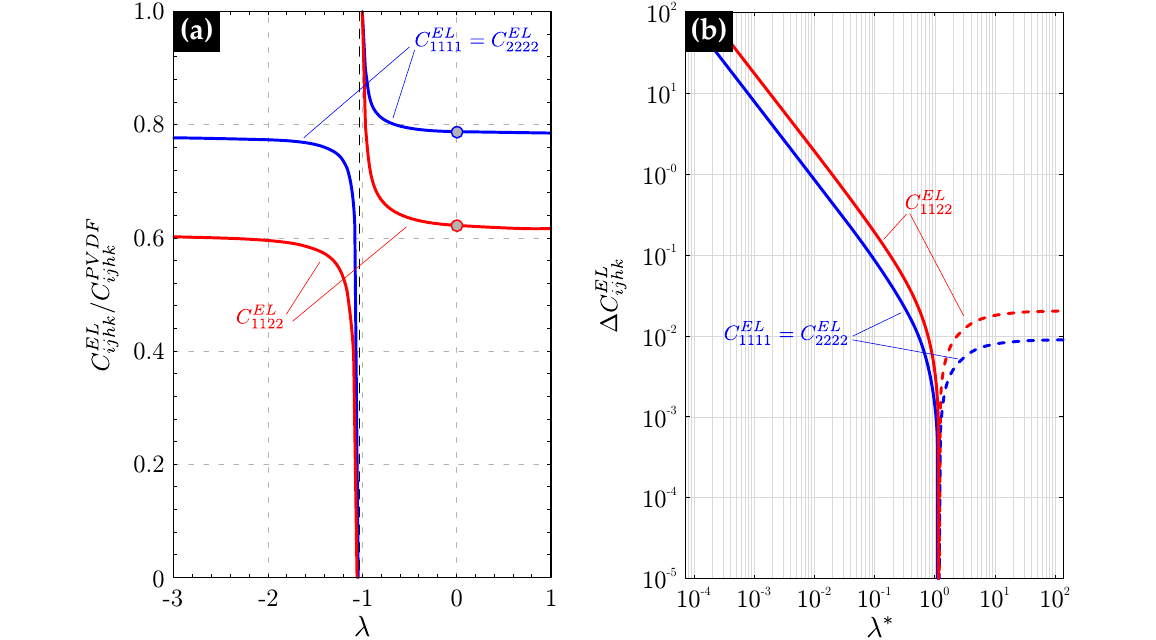}
    \caption{\footnotesize (a) Dimensionless components of the equivalent constitutive tensors $C^{EL}_{ijhk}/C^{PVDF}_{ijhk}$ versus the tuning parameter $\lambda$. (b) $\Delta C^{EL}_{ijhk}=  |C^{EL}_{ijhk}-C^{EL0}_{ijhk}|/C^{EL0}_{ijhk}$, versus the tuning variable $\lambda^*$ in log-log plot. The blue curves refer to the components $C^{EL}_{1111}=C^{EL}_{2222}$, while the red curves refer to the components $C^{EL}_{1122}$.  }
    \label{figure2n}
\end{figure}

We aim to define the time-dependent in-plane constitutive equations governing the shunting piezoelectric phase. In this regard, we start from a three-dimensional orthotropic piezoelectric material with polarization along the $\textbf{e}_3= \textbf{e}_1 \times \textbf{e}_2$ unit vector, according with the pioneering work by \cite{HAGOOD1991}, where each material point is identified by its position vector $\textbf{x} = x_i \textbf{e}_i $, with $i$=1,2,3, referred to a coordinate system with origin at point $O$ at time $t$. \textcolor{black}{The corresponding  coupled constitutive relations together with the related ones in the transformed Laplace space are detailed in Appendix A. After proper
 manipulations, the constitutive relations expressed in terms of 
the transformed electric potential difference $\mathord{\buildrel{\lower3pt\hbox{$\scriptscriptstyle\frown$}} \over V}_{3}$, imposed between the two opposite external surfaces of the piezoelectric ring orthogonal to $\textbf{e}_3$, and the transformed current $\mathord{\buildrel{\lower3pt\hbox{$\scriptscriptstyle\frown$}} \over I}_{3}$ along the $\textbf{e}_3$ direction are defined in the Laplace domain, become
\begin{align}
&\mathord{\buildrel{\lower3pt\hbox{$\scriptscriptstyle\frown$}} \over \sigma}_{ij}=C_{ijhk} \, \mathord{\buildrel{\lower3pt\hbox{$\scriptscriptstyle\frown$}} \over \varepsilon}_{hk}+C_{ij33} \, \mathord{\buildrel{\lower3pt\hbox{$\scriptscriptstyle\frown$}} \over \varepsilon}_{33}-\frac{e_{ij3} \mathord{\buildrel{\lower3pt\hbox{$\scriptscriptstyle\frown$}} \over V}_{3}}{L^{(P)}} ,\nonumber \\
&\mathord{\buildrel{\lower3pt\hbox{$\scriptscriptstyle\frown$}} \over \sigma}_{33}=C_{33hk} \, \mathord{\buildrel{\lower3pt\hbox{$\scriptscriptstyle\frown$}} \over \varepsilon}_{hk}+C_{3333} \, \mathord{\buildrel{\lower3pt\hbox{$\scriptscriptstyle\frown$}} \over \varepsilon}_{33}-\frac{e_{333} \mathord{\buildrel{\lower3pt\hbox{$\scriptscriptstyle\frown$}} \over V}_{3}}{L^{(P)}},\nonumber\\
&\mathord{\buildrel{\lower3pt\hbox{$\scriptscriptstyle\frown$}} \over \sigma}_{\alpha3}=2 \, C_{\alpha3\alpha3} \, \mathord{\buildrel{\lower3pt\hbox{$\scriptscriptstyle\frown$}} \over \varepsilon}_{\alpha3}+ e_{\alpha3\alpha} \,  \frac{\partial \mathord{\buildrel{\lower3pt\hbox{$\scriptscriptstyle\frown$}} \over \phi}}{\partial x_\alpha},\nonumber \\
&\mathord{\buildrel{\lower3pt\hbox{$\scriptscriptstyle\frown$}} \over D}_{i}=2 \, \widetilde{e}_{ij3} \, \mathord{\buildrel{\lower3pt\hbox{$\scriptscriptstyle\frown$}} \over \varepsilon}_{j3}+ \beta_{ij} \,  \frac{\partial \mathord{\buildrel{\lower3pt\hbox{$\scriptscriptstyle\frown$}} \over \phi}}{\partial x_j},\nonumber \\
&\mathord{\buildrel{\lower3pt\hbox{$\scriptscriptstyle\frown$}} \over I}_{3}= s \, A^{(P)}  \widetilde{e}_{3jh} \, \mathord{\buildrel{\lower3pt\hbox{$\scriptscriptstyle\frown$}} \over \varepsilon}_{jh}+ s \, A^{(P)} \, \widetilde{e}_{333} \, \mathord{\buildrel{\lower3pt\hbox{$\scriptscriptstyle\frown$}} \over \varepsilon}_{33}+  \, Y^{(P)}_{33}(s)  \, \mathord{\buildrel{\lower3pt\hbox{$\scriptscriptstyle\frown$}} \over V}_{3} , \qquad i,j,h,k,\alpha=1,2,\label{33}
\end{align}
where from now on no summation on index $\alpha$ is applied, and being $\mathord{\buildrel{\lower3pt\hbox{$\scriptscriptstyle\frown$}} \over \sigma}_{pq}$ the components of the transformed stress tensor, $\mathord{\buildrel{\lower3pt\hbox{$\scriptscriptstyle\frown$}} \over D}_{p}$ the components of the 
 transformed electric displacement field, $\mathord{\buildrel{\lower3pt\hbox{$\scriptscriptstyle\frown$}} \over \varepsilon}_{rs} = (\partial \mathord{\buildrel{\lower3pt\hbox{$\scriptscriptstyle\frown$}} \over u}_r/ \partial x_s+ \partial \mathord{\buildrel{\lower3pt\hbox{$\scriptscriptstyle\frown$}} \over u}_s /\partial x_r) / 2$ is the components of the  transformed strain tensor, with $\mathord{\buildrel{\lower3pt\hbox{$\scriptscriptstyle\frown$}} \over u}_p$ the transformed displacement field components, $\mathord{\buildrel{\lower3pt\hbox{$\scriptscriptstyle\frown$}} \over \phi}$ the transformed electric potential field, 
 $ C_{pqrs}$ the components of the fourth order elasticity tensor, 
$\beta_{pr}$
the components of the second order dielectric
permittivity tensor, ${e}_{qsp}$ the components of  the third order piezoelectric stress-charge coupling tensor and its transpose $\widetilde{e}_{pqs}={e}_{qsp}$. Moreover, the auxiliary variable   $Y^{(P)}_{33}(s)= \left(s A^{(P)} / L^{(P)} \right) \beta_{33}$ has been introduced, playing the role of the admittance of the piezoelectric material in the $\textbf{e}_3$ direction, being $ A^{(P)}= \pi ((R-h/2)^2-r^2)$ the area of the piezoelectric annulus, and $L^{(P)}=w$.}\\
In the case of shunting piezoelectric material, in which an electrical circuit   is connected in parallel to the piezoelectric ring, see Fig. \ref{figurePrima}(b), the equivalent shunting admittance $Y^{(SU)}_{33}$ sums to the piezoelectric one $Y^{(P)}_{33}$. Details concerning the constitutive relations of the electrical circuit in terms of transformed electric potential difference $\mathord{\buildrel{\lower3pt\hbox{$\scriptscriptstyle\frown$}} \over V}_{3}$ and of the transformed current $\mathord{\buildrel{\lower3pt\hbox{$\scriptscriptstyle\frown$}} \over I}_{3}$ are discussed in Appendix B. In agreement with this assumption, the last equation of (\ref{33}) specializes into 
\begin{align}
&\mathord{\buildrel{\lower3pt\hbox{$\scriptscriptstyle\frown$}} \over I}_{3}= s \, A^{(P)}  \widetilde{e}_{3jh} \, \mathord{\buildrel{\lower3pt\hbox{$\scriptscriptstyle\frown$}} \over \varepsilon}_{jh}+ s \, A^{(P)} \, \widetilde{e}_{333} \, \mathord{\buildrel{\lower3pt\hbox{$\scriptscriptstyle\frown$}} \over \varepsilon}_{33}+  \, Y^{(EL)}_{33}(s)  \, \mathord{\buildrel{\lower3pt\hbox{$\scriptscriptstyle\frown$}} \over V}_{3} , \qquad j,h=1,2,\label{340}
\end{align}
being $Y^{(EL)}_{33}(s)= Y^{(P)}_{33}(s)+ Y^{(SU)}_{33}(s)$ the total equivalent admittance. It is worth-noting that in the general case the constitutive variable $Y^{(EL)}_{33}$  explicitly depends  on the complex Laplace variable $s$. \\
After proper manipulations of equation (\ref{340}),
the electric potential difference $\mathord{\buildrel{\lower3pt\hbox{$\scriptscriptstyle\frown$}} \over V}_{3}$ is made explicit and can be plugged into (\ref{33}), so that
\begin{align}
&\mathord{\buildrel{\lower3pt\hbox{$\scriptscriptstyle\frown$}} \over \sigma}_{33}= \left(C_{33hk} + e_{333} \frac{s \, A^{(P)}}{L^{(P)}} Z^{(EL)}_{33}(s)\widetilde{e}_{3hk}\right)\mathord{\buildrel{\lower3pt\hbox{$\scriptscriptstyle\frown$}} \over \varepsilon}_{hk}   + \nonumber\\ & \, \, \, \, \, \, \, \,+\left(C_{3333}+ e_{333} \frac{s \, A^{(P)}}{L^{(P)}} Z^{(EL)}_{33}(s)\widetilde{e}_{333} \right) \mathord{\buildrel{\lower3pt\hbox{$\scriptscriptstyle\frown$}} \over \varepsilon}_{33}-\frac{e_{333} \,Z^{(EL)}_{33}(s)}{L^{(P)}} \mathord{\buildrel{\lower3pt\hbox{$\scriptscriptstyle\frown$}} \over I}_3,\label{35}
\end{align}
where the total equivalent impedance $Z^{(EL)}_{33}(s)=1 /Y^{(EL)}_{33}(s)$ is introduced. Moreover, the strain component $ \mathord{\buildrel{\lower3pt\hbox{$\scriptscriptstyle\frown$}} \over \varepsilon}_{33}$ can be made explicit in equation (\ref{35}) and plugged, in turn, into equation (\ref{33}), therefore the following in-plane relations are obtained
\begin{align}
&\mathord{\buildrel{\lower3pt\hbox{$\scriptscriptstyle\frown$}} \over \sigma}_{ij}= \left[ C_{ijhk}+ e_{ij3} \frac{s \, A^{(P)}}{L^{(P)}} Z^{(EL)}_{33}(s)\widetilde{e}_{3hk} - \left( C_{ij33}+ e_{ij3} \frac{s \, A^{(P)}}{L^{(P)}} Z^{(EL)}_{33}(s)\widetilde{e}_{333}\right) \times \right. \nonumber\\
&\, \, \, \, \, \, \, \, \, \, \,  \, \, \,\left. \times \frac{C_{33hk}+ e_{333} \frac{s \, A^{(P)}}{L^{(P)}} Z^{(EL)}_{33}(s)\widetilde{e}_{3hk}}{C_{3333}+ e_{333} \frac{s \, A^{(P)}}{L^{(P)}} Z^{(EL)}_{33}(s)\widetilde{e}_{333}}\right] \mathord{\buildrel{\lower3pt\hbox{$\scriptscriptstyle\frown$}} \over \varepsilon}_{hk}+ \nonumber\\ & \, \, \, \, \, \, \, \,-\left[ \frac{e_{ij3} \,Z^{(EL)}_{33}(s)}{L^{(P)}}-\left( C_{ij33}+ e_{ij3} \frac{s \, A^{(P)}}{L^{(P)}} Z^{(EL)}_{33}(s)\widetilde{e}_{333}\right)  \frac{\frac{e_{333} \,Z^{(EL)}_{33}(s)}{L^{(P)}} }{C_{3333}+ e_{333} \frac{s \, A^{(P)}}{L^{(P)}} Z^{(EL)}_{33}(s)\widetilde{e}_{333}} \right]\mathord{\buildrel{\lower3pt\hbox{$\scriptscriptstyle\frown$}} \over I}_3+ \nonumber\\
& \, \, \, \, \, \, \, \, \, \, \,  \, \, \,+ \frac{C_{ij33}+ e_{ij3} \frac{s \, A^{(P)}}{L^{(P)}} Z^{(EL)}_{33}(s)\widetilde{e}_{333}}{C_{3333}+ e_{333} \frac{s \, A^{(P)}}{L^{(P)}} Z^{(EL)}_{33}(s)\widetilde{e}_{333}}\mathord{\buildrel{\lower3pt\hbox{$\scriptscriptstyle\frown$}} \over \sigma}_{33} ,\nonumber \\
& \mathord{\buildrel{\lower3pt\hbox{$\scriptscriptstyle\frown$}} \over D}_{\alpha}= \frac{\widetilde{e}_{\alpha \alpha 3}}{C_{\alpha3\alpha3}} \mathord{\buildrel{\lower3pt\hbox{$\scriptscriptstyle\frown$}} \over \sigma}_{\alpha3}- \left( \beta_{\alpha \alpha} + \frac{ e_{\alpha 3 \alpha} \widetilde{e}_{\alpha \alpha 3}}{C_{\alpha 3 \alpha 3}}\right) \frac{\partial \mathord{\buildrel{\lower3pt\hbox{$\scriptscriptstyle\frown$}} \over \phi}}{\partial x_\alpha}, \qquad i,j,h,k,\alpha=1,2.\label{36}
\end{align}
For the sake of convenience, the total equivalent admittance of the shunting piezoelectric material can be expressed as $Y_{33}^{(EL)}(s)= s A^{(P)} \beta_{33}^{EL}(s) / L^{(P)} $,
where the auxiliary $s$-dependent function $\beta_{33}^{EL}(s)=\beta_{33}[1+L^{(P)} Y_{33}^{(SU)}(s) /(s\, \beta_{33} \, A^{(P)} )]= \beta_{33} \left( 1+ \lambda(s)\right)$ is introduced, with the \textcolor{black}{ so-called
 \textit{tuning function} $\lambda(s) = L^{(P)} Y_{33}^{SU}(s) /(s\, \beta_{33} \, A^{(P)} ) $ }linearly depending on the generic equivalent shunting admittance $Y_{33}^{(SU)}(s)$ that, in turn,  may depend on one or more tuning parameters characterizing the specific electrical circuit. \\
Focusing on the particular case characterized by 
$\mathord{\buildrel{\lower3pt\hbox{$\scriptscriptstyle\frown$}} \over I}_3=0$,  $\mathord{\buildrel{\lower3pt\hbox{$\scriptscriptstyle\frown$}} \over \sigma}_{i3}=0$ with $i=1,2,3$, 
the in-plane constitutive relations (\ref{36}) become
\begin{align}
&\mathord{\buildrel{\lower3pt\hbox{$\scriptscriptstyle\frown$}} \over \sigma}_{ij}= \left[ C_{ijhk}+ \frac{e_{ij3} \widetilde{e}_{3hk}}{\beta_{33}^{EL}(\lambda(s))}  - \left( C_{ij33}+\frac{e_{ij3} \widetilde{e}_{333}}{\beta_{33}^{EL}(\lambda(s))}\right)  \left( \frac{C_{33hk}+\frac{e_{333} \widetilde{e}_{3hk}}{\beta_{33}^{EL}(\lambda(s))}}{C_{3333}+ \frac{e_{333} \widetilde{e}_{333}}{\beta_{33}^{EL}(\lambda(s))}} \right) \right] \mathord{\buildrel{\lower3pt\hbox{$\scriptscriptstyle\frown$}} \over \varepsilon}_{hk}  ,\nonumber \\
& \mathord{\buildrel{\lower3pt\hbox{$\scriptscriptstyle\frown$}} \over D}_{\alpha}= - \left( \beta_{\alpha \alpha} + \frac{ e_{\alpha 3 \alpha} \widetilde{e}_{\alpha \alpha 3}}{C_{\alpha 3 \alpha 3}}\right) \frac{\partial \mathord{\buildrel{\lower3pt\hbox{$\scriptscriptstyle\frown$}} \over \phi}}{\partial x_{\alpha}}, \qquad i,j,h,k,\alpha=1,2.\label{37}
\end{align}
 Moreover, equations (\ref{37}) can be expressed in the compact form resulting in
\begin{align}
&\mathord{\buildrel{\lower3pt\hbox{$\scriptscriptstyle\frown$}} \over \sigma}_{ij}=  C^{EL}_{ijhk}(\lambda(s)) \, \mathord{\buildrel{\lower3pt\hbox{$\scriptscriptstyle\frown$}} \over \varepsilon}_{hk}  ,\nonumber \\
& \mathord{\buildrel{\lower3pt\hbox{$\scriptscriptstyle\frown$}} \over D}_{\alpha}= - \beta_{\alpha \alpha}^{EL}(\lambda(s)) \frac{\partial \mathord{\buildrel{\lower3pt\hbox{$\scriptscriptstyle\frown$}} \over \phi}}{\partial x_{\alpha}}, \qquad i,j,h,k,\alpha=1,2.\label{38}
\end{align}
It is worth-noting that the in-plane constitutive equations of an out-of-plane polarized piezoelectric continuum turn out to be uncoupled and, therefore, formally equivalent to the equations of a linear elastic dielectric material \citep{toupin1960stress}.
Thus, the components of the equivalent constitutive tensors pertaining to the shunting piezoelectric material result as
\begin{align}
&C^{EL}_{ijhk}(\lambda(s))= C_{ijhk}+ \frac{e_{ij3} \widetilde{e}_{3hk}}{\beta_{33}^{EL}(\lambda(s))}  - \left( C_{ij33}+\frac{e_{ij3} \widetilde{e}_{333}}{\beta_{33}^{EL}(\lambda(s))}\right)  \left( \frac{C_{33hk}+\frac{e_{333} \widetilde{e}_{3hk}}{\beta_{33}^{EL}(\lambda(s))}}{C_{3333}+ \frac{e_{333} \widetilde{e}_{333}}{\beta_{33}^{EL}(\lambda(s))}} \right)  ,\nonumber \\
& \beta_{\alpha \alpha}^{EL}= \beta_{\alpha \alpha} + \frac{ e_{\alpha 3 \alpha} \widetilde{e}_{\alpha \alpha 3}}{C_{\alpha 3 \alpha 3}}, \qquad i,j,h,k,\alpha=1,2,\label{39}
\end{align}
with the components $C^{EL}_{ijhk}$ explicitly depending on the tuning function $\lambda(s)$, except for the component $C^{EL}_{1212}$ that is independent on $\lambda(s)$. 
Note that, as expected, the equivalent elastic tensor satisfies the major and minor symmetries.\\\\
\textbf{\textit{Equivalent purely capacitive electrical circuit}}\\
\textcolor{black}{In the particular case of remarkable technological interest,  in which the equivalent electrical circuit is characterized by a purely capacitive equivalent admittance, it results  that the auxiliary function $\beta_{33}^{EL}(s)$ becomes $s$-independent, i.e. $\beta_{33}^{EL}=\beta_{33}+\mathcal{C} \frac{L^{(P)}}{A^{(P)}} = \beta_{33} \left( 1+ \lambda\right)$, being $Y_{33}^{SU}(s)= s \mathcal{C}$ and where $\lambda= (\mathcal{C} L^{(P)})/ (\beta_{33} A^{(P)})$, with $\lambda \in \mathbb{R}$, plays the role of a tuning parameter directly related to the capacitance $\mathcal{C}$.} Also negative values of $\lambda$ can be considered, exploiting
equivalent negative capacitance circuits as in
\cite{westra2000oscillators}. It follows that the equivalent elastic components $C^{EL}_{ijhk}$ in (\ref{39}) are, in turn, depend on the tuning parameter $\lambda$ and are $s$-independent.\\
In particular, in Fig. \ref{figure2n} the dependence of the elastic components  $C^{EL}_{ijhk}$ of the shunting piezoelectric material on the tuning variable is investigated in the case of the Polyvinylidene fluoride (PVDF), whose piezoelectric material properties are listed in the following Section \ref{Section4}. More specifically, in Fig. \ref{figure2n}(a) the dimensionless components $C^{EL}_{1111}/C^{PVDF}_{1111}= C^{EL}_{2222}/C^{PVDF}_{2222}$ (blue curves) and $C^{EL}_{1122}/C^{PVDF}_{1122}$ (red curves) are depicted versus $\lambda$. The curves exhibit vertical asymptotes in correspondence of the \textit{resonance} value of the tuning parameter $\lambda_R$, expressed in the following form
\begin{align}
&\lambda_R= -\left( 1+ \frac{e_{333}}{C_{3333} \beta_{33}}\right)  . \label{310}
\end{align}
\noindent \textcolor{black}{It is worth noting that the resonance phenomenon occurs for a negative $\lambda$ value corresponding to a negative purely capacitive equivalent admittance $\mathcal{C}$. }
\noindent Further horizontal asymptotes are approached as 
$\lambda\to \pm \infty $, corresponding to the following asymptotic values 
\begin{align}
& C^{EL \infty}_{ijhk}= \lim_{\lambda\to \pm \infty} C^{EL}_{ijhk}(\lambda)=C_{ijhk}-\frac{C_{ij33} C_{hk33}}{C_{3333}}, \qquad i,j,h,k=1,2. \label{311}
\end{align}
\textcolor{black}{ Finally, for $\lambda$=0 the equivalent elastic components  $C^{EL0}_{ijhk}$, see highlighted points in Fig. \ref{figure2n}(a), turn to be those of 
a standard piezoelectric material coupled with an equivalent electrical circuit characterized by zero capacitance (or also an open circuit case), resulting in}
\begin{align}
& C^{EL0}_{ijhk}=C^{EL}_{ijhk}(\lambda=0)=C_{ijhk}+\frac{e_{ij3}e_{hk3}}{\beta_{33}}-\frac{\left(C_{ij33}+\frac{e_{ij3} e_{333}}{\beta_{33}}\right)\left(C_{hk33}+\frac{e_{hk3} e_{333}}{\beta_{33}}\right)}{C_{3333}+\frac{(e_{333})^2}{\beta_{33}}}, \quad i,j,h,k=1,2. \label{310}
\end{align}
\textcolor{black}{Figure \ref{figure2n}(b) shows, in log-log plot, the absolute value of
 the relative difference between the component $C^{EL}_{ijhk}$ and the corresponding one evaluated in $\lambda$=0, i.e. $\Delta C^{EL}_{ijhk}= |C^{EL}_{ijhk}-C^{EL0}_{ijhk}|/C^{EL0}_{ijhk}$, as a function of the
shifted tuning variable $\lambda^*= \lambda-\lambda_R$, investigated for $\lambda \ge \lambda_R$.}  In particular, the positive values of $C^{EL}_{ijhk}-C^{EL0}_{ijhk}$ are plotted in
solid lines, while the negative ones in dashed lines. As expected, an asymptotic behaviour is found for $\lambda^*$=1. Furthermore, a stiffer behaviour is exhibited as $\lambda^*$ approaches zero, while a softer behaviour is shown for $\lambda^* \to  \infty $. In this context, it is shown that by adjusting the equivalent impedance of the electrical circuit, it is possible to properly tuning the constitutive properties of the piezoelectric shunting element. It follows that the acoustic behaviour of the active metamaterial can be suitably tuned in turn.\\
\subsection{Governing equations of  tunable metamaterials} \label{Section22}
\begin{figure}[t]
	\includegraphics[width=1\textwidth]{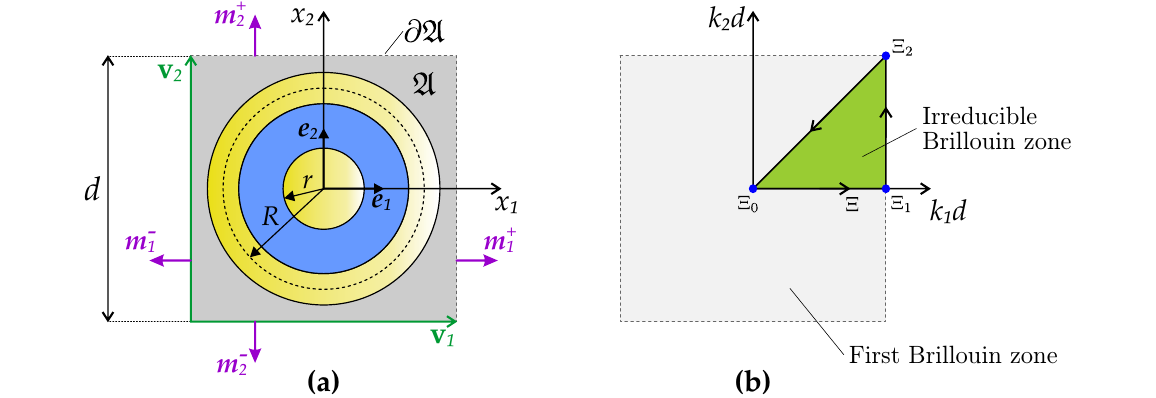}
	\caption{\footnotesize (a) Periodic cell together with its periodicity vectors $\textbf{v}_1$, $\textbf{v}_2$; (b) First Brillouin zone
		and the irreducible Brillouin zone highlighted in  green. }
	\label{figure2d}
\end{figure}

We aim to investigating the acoustic wave propagation in the tunable metamaterial.
In this regard, let us consider the periodic three-phase microstructured metamaterial in Fig \ref{figurePrima}, with a  piezoelectric phase shunted by a generic dissipative electrical circuit, described as a first order continuum in the framework of the in-plane linear theory.\\
The in-plane dynamic balance equations, in the transformed Laplace space, of the tunable infinite metamaterial are expressed in the following form 
\begin{align}\label{221}
\begin{split}
&\frac{\partial \mathord{\buildrel{\lower3pt\hbox{$\scriptscriptstyle\frown$}} \over \sigma_{ij}}}{\partial x_j} + \mathord{\buildrel{\lower3pt\hbox{$\scriptscriptstyle\frown$}} \over b_i} =\rho  s^2 \mathord{\buildrel{\lower3pt\hbox{$\scriptscriptstyle\frown$}} \over u_i},  \quad  \qquad i,j=1,2,
\end{split}
\end{align}
\textcolor{black}{being $\mathord{\buildrel{\lower3pt\hbox{$\scriptscriptstyle\frown$}} \over \sigma_{ij}}$ and $\mathord{\buildrel{\lower3pt\hbox{$\scriptscriptstyle\frown$}} \over u_i}$ the in-plane stress components and the in-plane displacement components, respectively, $\mathord{\buildrel{\lower3pt\hbox{$\scriptscriptstyle\frown$}} \over b_i}$ the transformed source term (or body force) and $\rho$ the mass density and $x_j$ the components of the in-plane position vector $\textbf{x} = x_j \textbf{e}_j $, with $j$=1, 2.}\\
By exploiting the constitutive relations 
\begin{align}\label{222}
\begin{split}
& \mathord{\buildrel{\lower3pt\hbox{$\scriptscriptstyle\frown$}} \over \sigma_{ij}}=  C^\textrm{\ding{71}}_{ijkl}(s) \frac{\partial \mathord{\buildrel{\lower3pt\hbox{$\scriptscriptstyle\frown$}} \over u_k} }{\partial x_l} ,  \quad  \qquad i,j,k,l=1,2,
\end{split}
\end{align}
where the components $C^\textrm{\ding{71}}_{ijkl}$ of the constitutive tensor for the shunting piezoelectric phase
are in general $s$-dependent and coincide with $C^{EL}_{ijkl}$ expressed in (\ref{39}), while for the other elastic phases are $s$-independent. \\
Because of the periodicity of the metamaterial, it is worth noting that both constitutive tensors and the mass density satisfy the $\mathfrak{A}$ periodicity, i.e.
\begin{align}
&C^\textrm{\ding{71}}_{ijkl}\left( {{\textbf{x}} + {{\mathbf{v}}_{\beta}}},s \right) = C^\textrm{\ding{71}}_{ijkl}\left( {\textbf{x}},s \right),\nonumber \\
&\rho\left( {{\textbf{x}} + {{\mathbf{v}}_{\beta}}} \right) =\rho\left( {\textbf{x}} \right), \qquad i,j,k,l=1,2,  \qquad \beta=1,2, \quad\forall \textbf{x} \in \mathfrak{A},\label{22}
\end{align}
being ${\mathbf{v}}_{\beta}$ the periodicity vector.\\
By plugging (\ref{222}) in (\ref{221}), the governing equation of the in-plane free Bloch wave propagation, i.e. with $\mathord{\buildrel{\lower3pt\hbox{$\scriptscriptstyle\frown$}} \over b_i}=0$, in the transformed Laplace space results
\begin{align}\label{241}
\begin{split}
&\frac{\partial}{\partial x_j} \left(  C^\textrm{\ding{71}}_{ijkl}(s) \frac{\partial \mathord{\buildrel{\lower3pt\hbox{$\scriptscriptstyle\frown$}} \over u_k} }{\partial x_l} \right)-\rho  s^2 \mathord{\buildrel{\lower3pt\hbox{$\scriptscriptstyle\frown$}} \over u_i}=0,  \quad  \qquad i,j,k,l=1,2.
\end{split}
\end{align}
By exploiting the medium periodicity, it is possible to consider only a single Periodic Cell $\mathfrak{A}$ undergoing the following Floquet-Bloch boundary conditions
\begin{align}
&\mathord{\buildrel{\lower3pt\hbox{$\scriptscriptstyle\frown$}} \over u_i^+}   = \mathord{\buildrel{\lower3pt\hbox{$\scriptscriptstyle\frown$}} \over u_i^-}  {e^{\iota k_j v_j^{(\beta)}}}, \nonumber\\
& \mathord{\buildrel{\lower3pt\hbox{$\scriptscriptstyle\frown$}} \over {\sigma}_{lr}^{+}}  =-  \mathord{\buildrel{\lower3pt\hbox{$\scriptscriptstyle\frown$}} \over {\sigma}_{lr}^{-}}  (m_r^{(\beta)})^- e^{\iota k_j v_j^{(\beta)}},\label{25}
\end{align}
where $v_j^{(\beta)}$ are the components of the periodicity vector 
$\textbf{v}_{\beta}= v_j^{(\beta)} \textbf{e}_j$,  and $(m_r^{(\beta)})^{\pm}$ are the components of the outward normal $\textbf{m}_{\beta}^{\pm}= (m_j^{(\beta)})^{\pm} \textbf{e}_j$ to the boundary $\partial \mathfrak{A}$, $j,\beta=1,2$, see Fig. \ref{figure2d}(a). In addition, the superscripts $^{\pm}$ 
refer to the positive part  $\partial \mathfrak{A}^{+}$ (with outward normal $\textbf{m}_{\beta}^{+}$) and  to the corresponding negative parts  $\partial \mathfrak{A}^{-}$ (with outward normal $\textbf{m}_{\beta}^{-}$) of the Periodic Cell boundary. \textcolor{black}{We recall that the following notation applies to the generic function $\mathord{\buildrel{\lower3pt\hbox{$\scriptscriptstyle\frown$}} \over g}$, i.e.
$\mathord{\buildrel{\lower3pt\hbox{$\scriptscriptstyle\frown$}} \over g}^{\pm} 
\buildrel\textstyle.\over=\mathord{\buildrel{\lower3pt\hbox{$\scriptscriptstyle\frown$}} \over g}(\textbf{x}^{\pm}) $
where $\textbf{x}^{\pm}$ identifies the generic point $P$ $ \in \partial \mathfrak{A}^{\pm}$, and  $\textbf{x}^{+}=\textbf{x}^{-}+\textbf{v}_{\beta}$.} Furthermore, $k_j$ are the components of the wave vector $\textbf{k}$.
The dimensionless first Brillouin zone 
$\mathfrak{B}=[-\pi, \pi] \times [-\pi, \pi] $ is defined in the space of dimensionless wave vectors and is characterized by  two orthogonal vectors $\pi \textbf{n}_i$, parallel to $\textbf{e}_i$, with $i$=1, 2, coinciding with the periodicity 
vectors by virtue of the symmetry properties of the micro-structure, see Fig. \ref{figure2d}(b). The solution of the differential eigen problem (\ref{241}) with (\ref{25}) leads in general to a complex-value frequency band structure, accordingly with the pioneering Floquet-Bloch theory \citep{Floquet1883,Bloch1928,Brillouin1960}. \\
Note that in the particular case of periodic material including a piezoelectric phase shunted by a non-dissipative electrical circuit or characterized by an equivalent purely capacitive electrical circuit, it can be proven that the 
complex frequency $s$ turns to be
purely imaginary, i.e. $s=\iota \omega$, being $\omega \in \mathbb{R}$ the angular frequency, and $\iota^2=-1$. 
More specifically, in this case the differential eigen problem (\ref{241}) takes the form
\begin{align}\label{26}
\begin{split}
&\frac{\partial}{\partial x_j} \left(  C^\textrm{\ding{71}}_{ijkl}\frac{\partial \mathord{\buildrel{\lower3pt\hbox{$\scriptscriptstyle\frown$}} \over u_k} }{\partial x_l} \right)+\rho  \omega^2 \mathord{\buildrel{\lower3pt\hbox{$\scriptscriptstyle\frown$}} \over u_i}=0,  \quad  \qquad i,j,k,l=1,2  .
\end{split}
\end{align}
together with the same Floquet-Bloch boundary conditions as in (\ref{25}).\\
For the sake of completeness, the time domain governing equation of the in-plane free Bloch wave propagation in the periodic tunable metamaterial, shunted by a generic dissipative electrical circuit, can be determined by applying to (\ref{241}) the inverse bilateral Laplace transform 
\begin{align} \label{251}
\mathcal{L}^{-1}  \left[\mathord{\buildrel{\lower3pt\hbox{$\scriptscriptstyle\frown$}} \over g} (\textbf{x},s) \right]= \frac{1}{2 \pi \iota} \int_{r -\iota \infty}^{r +\iota \infty}\mathord{\buildrel{\lower3pt\hbox{$\scriptscriptstyle\frown$}} \over g} (\textbf{x},s)e^{s t}d{s}, \qquad r \in \mathbb{R}.
\end{align}
It results that the following integro-differential equation is obtained
\begin{align}\label{27}
\begin{split}
& \frac{\partial}{\partial x_j}\left( \mathcal{L}^{-1}  \left[ \frac{ C^\textrm{\ding{71}}_{ijkl}(s)}{s}  \right] * \frac{\partial^2 u_k }{\partial x_l \, \partial t} \right)=\rho  \frac{\partial^2 u_i}{\partial t^2}  \quad  \qquad i,j,k,l=1,2  .
\end{split}
\end{align}
where $*$ is the convolution product, and the well-known property $\mathcal{L}(\partial^n g(\textbf{x},t) / \partial t^n)=s^n  \mathord{\buildrel{\lower3pt\hbox{$\scriptscriptstyle\frown$}} \over g}(\textbf{x},s)$ is exploited. Note that such equation is formally analogous to the field equation of a periodic viscoelastic continuum, where the relaxation function $ C^\textrm{\ding{71}}_{ijkl}(s)/s $ explicitly depends on the equivalent admittance of a generic electrical circuit.\\
In the particular case of purely capacitive equivalent admittance, where the constitutive equivalent components $ C^\textrm{\ding{71}}_{ijkl} $ are $s$-independent, by applying to (\ref{26}) the inverse Fourier transform
\begin{align} \label{231}
\mathcal{F}^{-1}  \left[\mathord{\buildrel{\lower3pt\hbox{$\scriptscriptstyle\frown$}} \over g} (\textbf{x},\omega) \right]= \frac{1}{2 \pi } \int_{ - \infty}^{+\infty}\mathord{\buildrel{\lower3pt\hbox{$\scriptscriptstyle\frown$}} \over g} (\textbf{x},\omega)e^{\iota \omega t}d{\omega}, 
\end{align}
the field equation in the time domain takes the following form
\begin{align}\label{261}
\begin{split}
&\frac{\partial}{\partial x_j} \left(  C^\textrm{\ding{71}}_{ijkl}\frac{\partial u_k }{\partial x_l} \right)= \rho  \frac{ \partial^2 u_i}{\partial t^2},  \quad  \qquad i,j,k,l=1,2,
\end{split}
\end{align}
that is formally analogous to the field equation of
a periodic elastic first order continuum.

\textcolor{black}{\section{Numerical experiments}
\label{Section4}
This Section is first devoted to characterize the frequency band structure of the proposed tunable acoustic metamaterial for different geometric design parameters (Section \ref{Section31}). Furthermore, 
a numerical experiment is proposed to test the effectiveness of the periodic tunable metamaterial as acoustic filter
(Section \ref{Section41}).  
  }

\textcolor{black}{\subsection{Tuning of pass and stop bands}\label{Section31}
Here we investigate the in-plane dynamic response of the tunable acoustic metamaterial reported in Fig. \ref{figurePrima}(b), where also the relevant dimensions are depicted. } The radius of the resonator and the mean radius of the external ring are denoted by $r$ and $R$, respectively, while the edge of the square periodic cell is $d$. We assume that the soft matrix is made of Epotex 301, a passive polymer materials, with elastic properties $E$= 3.6 GPa and $\nu$=0.35,  and mass density $\rho$=1150 kg/m$^3$, see \citep{lee2014high}. Both the outer ring and the circular resonator are made of 
steel with $E$=210 GPa, $\nu$=0.3 and mass density $\rho$= 7500 kg/m$^3$. Focusing on the state of plane stress of such two materials, the non-vanishing components of the  elasticity tensor are determined as $C^\textrm{\ding{71}}_{1111}=C^\textrm{\ding{71}}_{2222}=E/(1-\nu^2)$, 
$C^\textrm{\ding{71}}_{1122}=\nu E/(1-\nu^2)$, $C^\textrm{\ding{71}}_{1212}= E/2(1+\nu)$. Finally, the shunted piezoelectric element is designed by connecting  a Polyvinylidene fluoride (PVDF, mass density $\rho$= 1780 kg/m$^3$) ring to an electrical circuit characterized by purely capacitive equivalent admittance, as schematically shown in Fig. \ref{figurePrima}(b). In this way, by modifying 
the tuning parameter $\lambda$ (introduced in Section \ref{Section3}), directly related to the equivalent impedance of the electrical circuit, it is possible to 
properly adjust the equivalent constitutive properties of the shunted piezoelectric element. 
 The electromechanical properties of PVDF, polarized along the $\textbf{e}_3$ direction, are taken from \citep{IYER2014} and are listed below.
The non vanishing components of the  elasticity tensor are $C_{1111}$ = $C_{2222}$ = 4.84 $\cdot 10^{09}$ Pa, 
$C_{3333}$ = 4.63$\cdot 10^{9}$ Pa,
$C_{1122}$ = 2.72$\cdot 10^{9}$ Pa,
$C_{1133}$ = $C_{2233}$ = 2.22$\cdot 10^{9}$ Pa,
$C_{1212}$=1.06$\cdot 10^{9}$ Pa, 
$C_{1313}$ = $C_{2323}$ = 5.26$\cdot 10^{7}$ Pa.
The non vanishing components of the stress-charge coupling tensor are  
$e_{113}$=$e_{223}$=-1.999 $\cdot 10^{-3}$ C/$\text{m}^2$, 
$e_{311}$=$e_{322}$=4.344 $\cdot 10^{-3}$ C/$\text{m}^2$, $e_{333}$=-1.099 $\cdot 10^{-1}$ C/$\text{m}^2$.
The set of components is complemented by the non vanishing components of the dielectric permittivity tensor, i.e. $\beta_{11}$=$\beta_{22} $= 6.641 $\cdot 10^{-11}$ C/Vm, 
 and $\beta_{33}$=7.083 $\cdot 10^{-11}$ C/Vm.
The corresponding in-plane constitutive components of the shunted piezoelectric material, characterized by $\mathord{\buildrel{\lower3pt\hbox{$\scriptscriptstyle\frown$}} \over I}_3=0$,  $\mathord{\buildrel{\lower3pt\hbox{$\scriptscriptstyle\frown$}} \over \sigma}_{i3}=0$ with $i=1,2,3$, are obtained by exploiting the equations (\ref{39}) in Section \ref{Section3}, where $C_{ijhk}$ =$C^{EL}_{ijhk}$, and  
$\beta_{ij}$=$\beta^{EL}_{ij}$, with $i,j,h,k=1,2$,  \textcolor{black}{ specialized to the case of tuning function $\lambda(s)=(\mathcal{C} L^{(P)})/ (\beta_{33} A^{(P)})$}. \\
We aim to study the influence of a set of geometric parameters, characterizing the periodic cell, on the Floquet-Bloch spectrum of the metamaterial as the tuning parameter $\lambda$ changes. In this regard, different periodic cells are taken into account, which differs for the ratios $r/R$ and  $R/d$, taking  the ratios $h/d$=0.075,  $w/d$=1, with the edge of the periodic cell $d$=1 cm. 
For the sake of clarity a simple schematic of the geometry together with the relevant dimensions is depicted in Figure \ref{figure2d}(a).
\textcolor{black}{Attention is focused on the acoustic band structure of the metamaterial in the first irriducible Brillouin zone (highlighted in green in Figure \ref{figure2d}(b)) defined within the closed polygonal curve $\Gamma$, joining the vertices $\Xi_0$, $\Xi_1$ and $\Xi_2$ placed on the curvilinear abscissa $\Xi$ in the dimensionless plane
($k_1 \, d$, $k_2 \, d$).}
\textcolor{black}{To this aim the governing equation of the in-plane free Bloch wave propagation (\ref{26}) together with the associated Floquet-Bloch boundary conditions (\ref{25}) are solved in the periodic cell exploiting the finite element method.  Triangular elements with quadratic Lagrangian interpolation functions have been adopted.  } Note that due to the centro-symmetry of the periodic cell, the curve $\Gamma$ delimits the first irreducible Brillouin zone.
\begin{figure}[H]
	\includegraphics[width=1\textwidth]{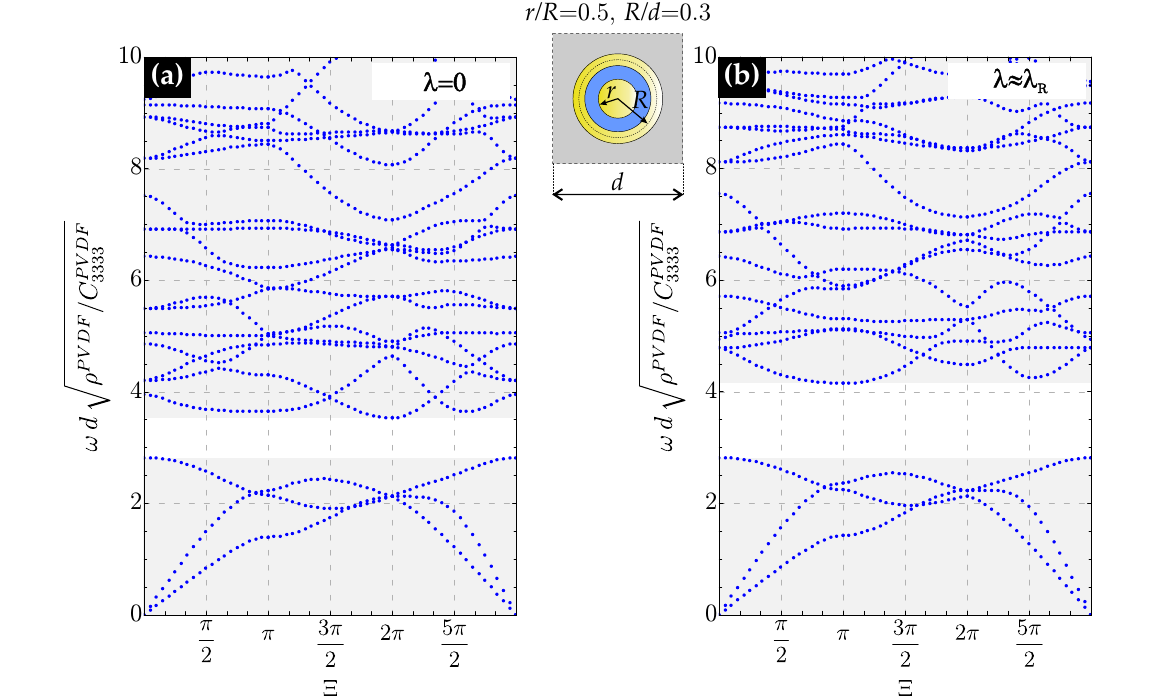}
	\caption{\footnotesize Floquet-Bloch spectrum for the metamaterial with $r/R$=0.5 and $R/d$=0.3 in terms of the dimensionless frequency $\omega \, d \sqrt{\rho^{PVDF}}/\sqrt{C^{PVDF}_{3333}}$ versus the curvilinear abscissa $\Xi$. (a) Tuning parameter $\lambda$=0; (b) tuning parameter $\lambda  \approx \lambda_R$.}
	\label{figure4}
\end{figure}
 \noindent First, three cases are considered characterized by 
$r/R$=0.5 and variable $R/d$. In Figures \ref{figure4}, \ref{figure5} and \ref{figure6}, the ratio $R/d$ takes values 0.3, 0.35 and 0.4, respectively. 
\begin{figure}[h]
	\includegraphics[width=1\textwidth]{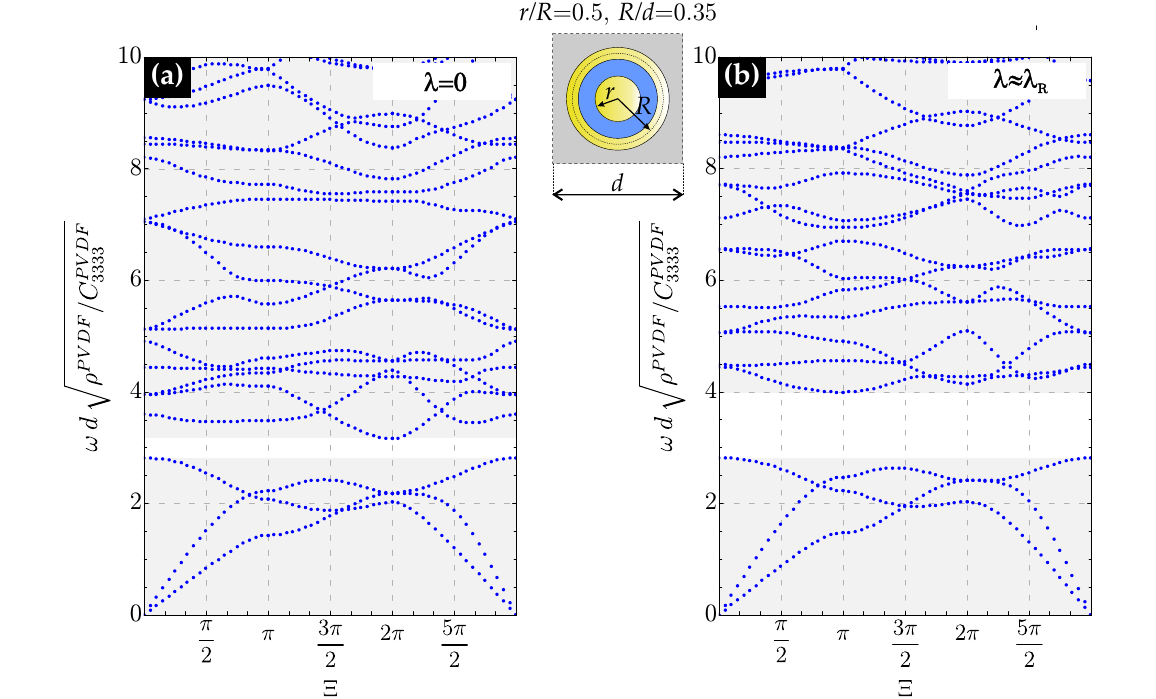}
	\caption{\footnotesize Floquet-Bloch spectrum for the metamaterial with $r/R$=0.5 and $R/d$=0.35 in terms of the dimensionless frequency $\omega \, d \sqrt{\rho^{PVDF}}/\sqrt{C^{PVDF}_{3333}}$ versus the curvilinear abscissa $\Xi$. (a) Tuning parameter $\lambda$=0; (b) tuning parameter $\lambda  \approx \lambda_R$.}
	\label{figure5}
\end{figure}
In all cases the comparison of the Floquet-Bloch spectrum obtained for $\lambda$=0, in figures (a), and $\lambda  \approx \lambda_R \simeq -1.0368$ , in figures (b),
is performed. The curves report the values of the dimensionless angular frequency $\omega \, d \sqrt{\rho^{PVDF}}/\sqrt{C^{PVDF}_{3333}}$,  against the curvilinear abscissa $\Xi$. The first 20 branches of
the spectrum are taken into account.
More specifically, the case with the smallest resonator ($R/d$=0.3) is shown in Figure \ref{figure4}. 
By comparing the results obtained for $\lambda$=0, Figure \ref{figure4}(a), i.e. the case of standard (non shunted) piezoelectric ring, with those for $\lambda  \approx \lambda_R$ i.e. $\textit{resonant}$ value, Figure \ref{figure4}(b), it emerges that the total low-frequency band gap, between the 1$^{st}$ and the 2$^{nd}$ optical branches, becomes noticeably wider. Moreover,  the two acoustic branches and the 1$^{st}$ optical branch remain almost unchanged as $\lambda$ varies, while in the considered frequency range a higher spectral density is detected for the remaining optical branches.\\
In Figure \ref{figure5} the case with $R/d$=0.35 is taken into account. In this case for $\lambda$=0, Fig. \ref{figure5}(a), three total band gaps are found. Indeed, besides the wider low frequency band gap, between the first and the second optical branches, two more narrow band gaps appear at higher frequencies in the considered range (between the 6$^{th}$ and the 7$^{th}$ optical branches and between the 11$^{th}$ and the 12$^{th}$ optical branches, respectively). Passing to $\lambda  \approx \lambda_R$, in Fig. \ref{figure5}(b), it is noticed that the low frequency band gap opens, the two further total band gaps disappear, while a new total band gap 
forms between the 8$^{th}$ and the 9$^{th}$ optical branches.\\
Finally, in Figure \ref{figure6} the ratio $R/d$=0.4 is considered. By comparing the Floquet-Bloch spectra for $\lambda$=0 and $\lambda  \approx \lambda_R$, it stands to reason that in the first case, Fig. \ref{figure6}(a), a very dense spectrum is found, with acoustic optical branches intersecting
each other over and over, without total band gaps. On the other hand, in the second case, Fig. \ref{figure6}(b), a wide low frequency band gap is detected, together with three further total band gaps opening between optical branches 6$^{th}$-7$^{th}$, 12$^{th}$-13$^{th}$ and 13$^{th}$-14$^{th}$, respectively.
\newpage
\begin{figure}[H]
	\includegraphics[width=1\textwidth]{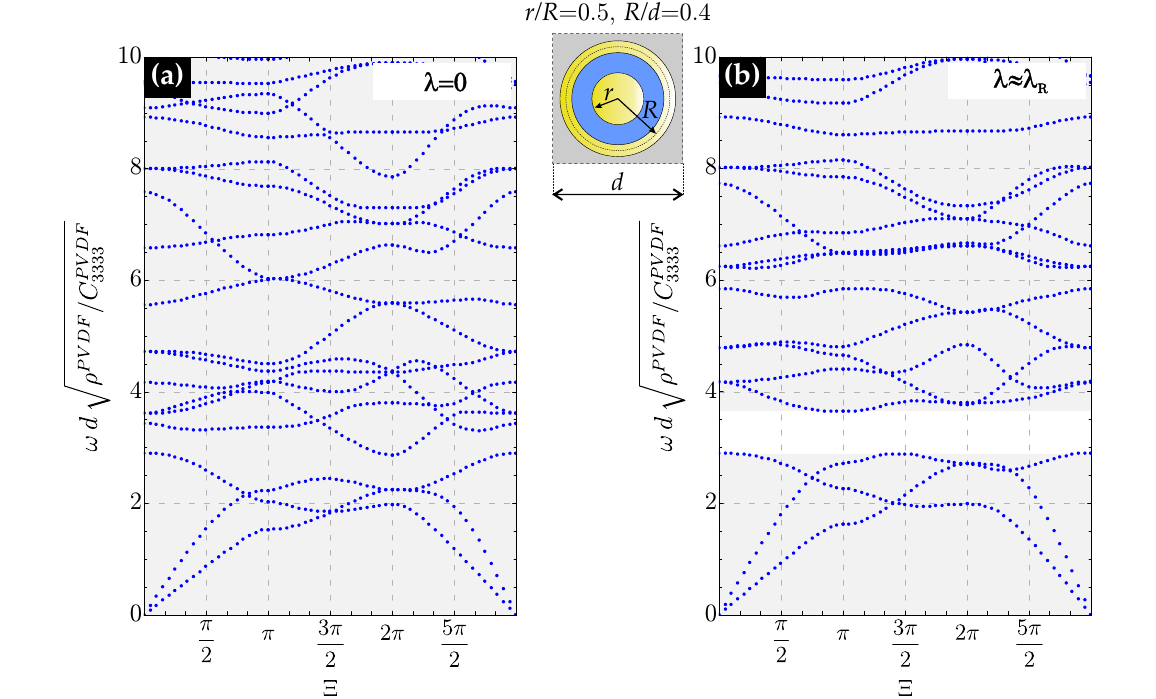}
	\caption{\footnotesize Floquet-Bloch spectrum for the metamaterial with $r/R$=0.5 and $R/d$=0.4 in terms of the dimensionless frequency $\omega \, d \sqrt{\rho^{PVDF}}/\sqrt{C^{PVDF}_{3333}}$ versus the curvilinear abscissa $\Xi$. (a) Tuning parameter $\lambda$=0; (b) tuning parameter $\lambda  \approx \lambda_R$.}
	\label{figure6}
\end{figure}

\begin{figure}[h]
	\includegraphics[width=1\textwidth]{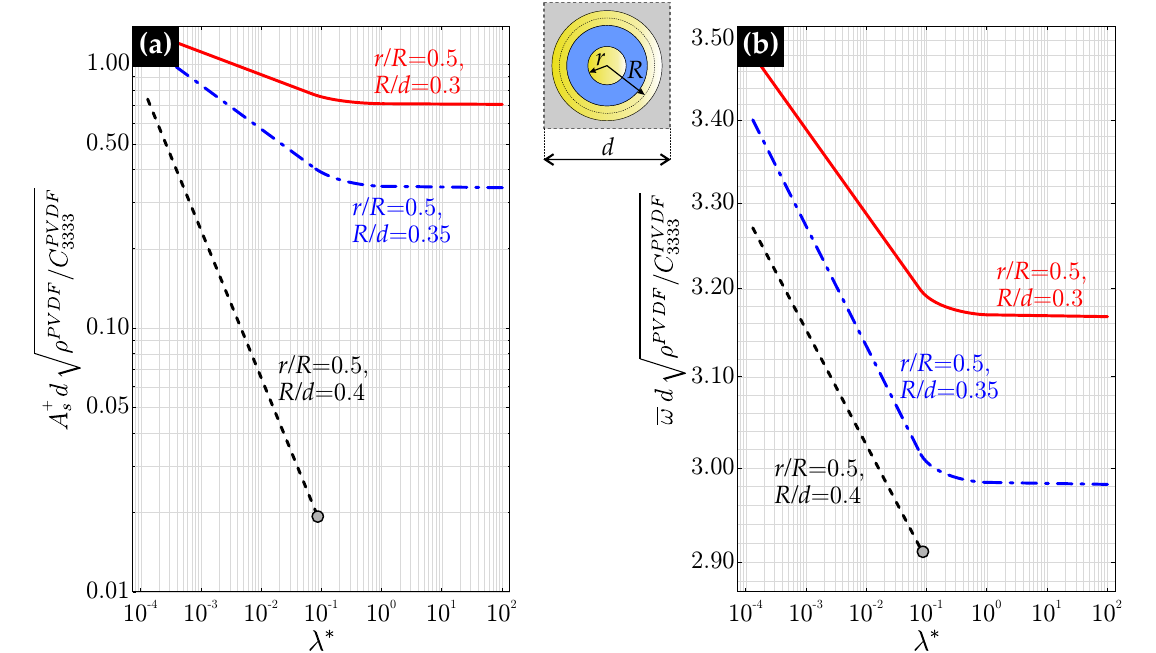}
	\caption{ \footnotesize Behaviour of the metamaterial with $r/R$=0.5 and $R/d$=0.3 (blue curves); $R/d$=0.35 (red curves) and $R/d$=0.4 (black curves). (a) Dimensionless band gap amplitude $A_s^+ d \sqrt{\rho^{PVDF}}/\sqrt{C_{3333}^{PVDF}}$ versus the tuning parameter $\lambda^*=\lambda-\lambda_R$; (b) Central frequency $\overline \omega d \sqrt{\rho^{PVDF}}/\sqrt{C_{3333}^{PVDF}}$ versus the tuning parameter $\lambda^*=\lambda-\lambda_R$. }
	\label{figure7}
\end{figure}
\begin{figure}[H]
	\includegraphics[width=1\textwidth]{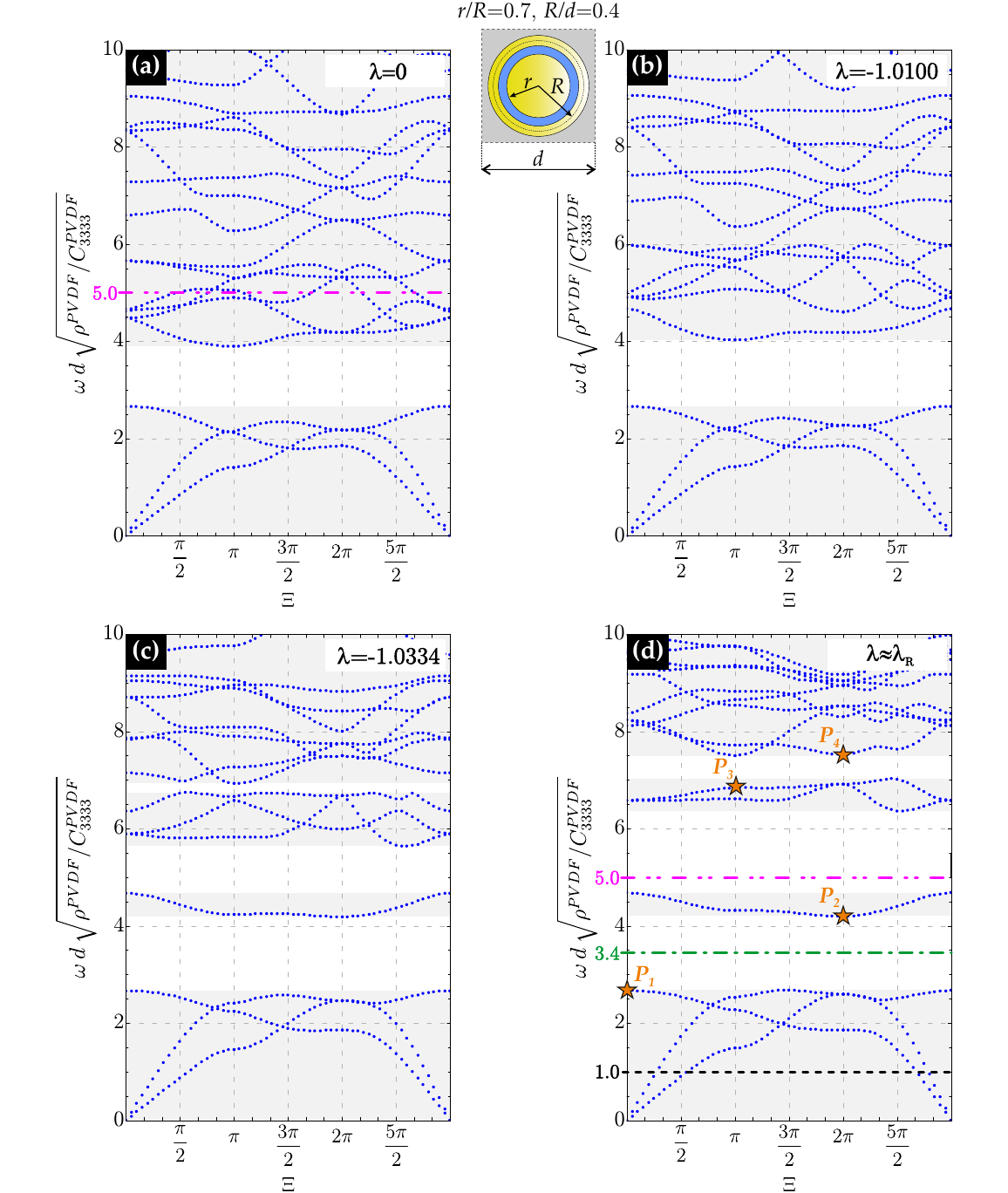}
	\caption{\footnotesize \textcolor{black}{Floquet-Bloch spectrum for the metamaterial with $r/R$=0.7 and $R/d$=0.4 in terms of the dimensionless frequency $\omega \, d \sqrt{\rho^{PVDF}}/\sqrt{C^{PVDF}_{3333}}$ versus the curvilinear abscissa $\Xi$. (a) Tuning parameter $\lambda$=0; (b) tuning parameter $\lambda$=-1.0100; (c) tuning parameter $\lambda$=-1.0334; (d) tuning parameter $\lambda  \approx \lambda_R$. In the subfigure (d), are also reported the dimensionless excitation frequencies ($\Omega/\Omega_r$=1 and 3.4)  exploited in the numerical experiments reported in Section \ref{Section41}.}}
	\label{figure8}
\end{figure}
\begin{figure}[H]
	\includegraphics[width=1\textwidth]{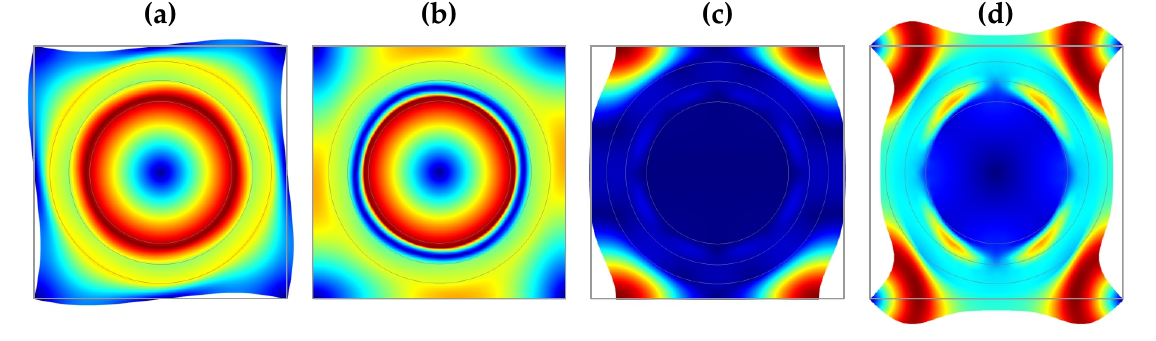} 
	\caption{\textcolor{black}{\footnotesize  Wave forms related to some relevant points of the Floquet-Bloch spectrum for the metamaterial with $r/R$=0.7 and $R/d$=0.4 for $\lambda  \approx \lambda_R$ in Figure \ref{figure8}(d) (a) Point $P_1$ in the 1$^{st}$ optical branch; (b) Point $P_2$ in the 2$^{nd}$ optical branch; (c) Point $P_3$ in the 4$^{th}$ optical branch; (d) Point $P_4$ in the 5$^{th}$ optical branch.}}
	\label{figure8a}
\end{figure}
\begin{figure}[H]
  \includegraphics[width=1\textwidth]{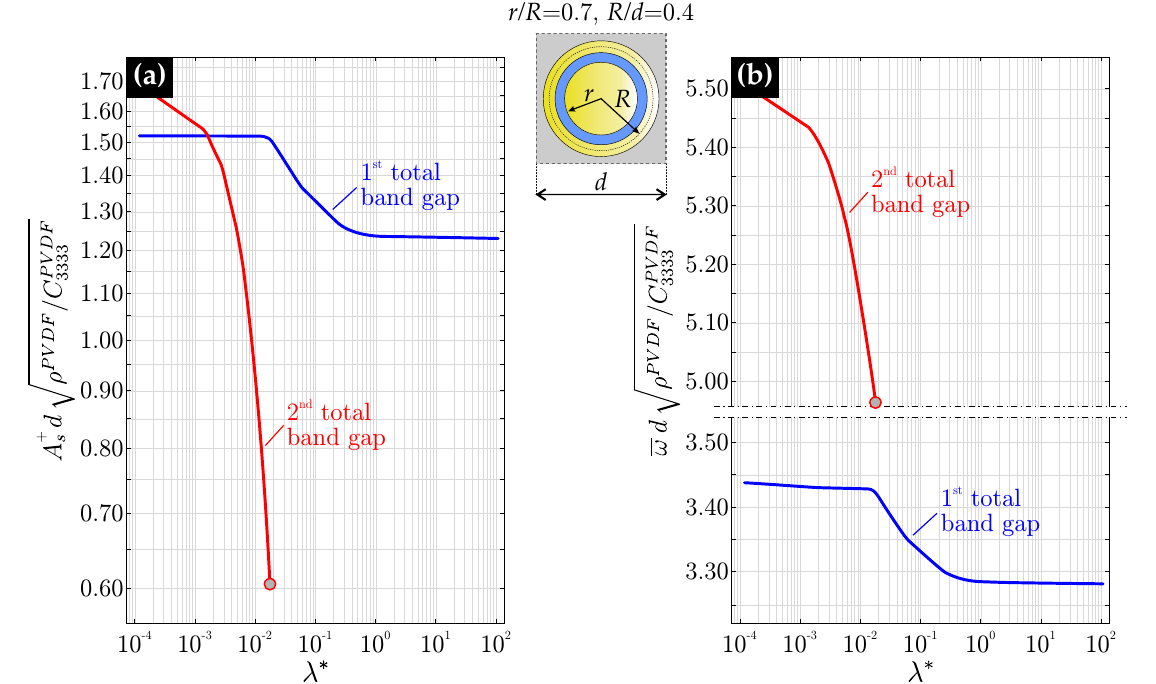}
    \caption{ \footnotesize First (blue curve) and second (red curve) total band gaps for the metamaterial with $r/R$=0.7 and $R/d$=0.4. (a) Dimensionless band gap amplitude $A_s^+ d \sqrt{\rho^{PVDF}}/\sqrt{C_{3333}^{PVDF}}$ versus the tuning parameter $\lambda^*=\lambda-\lambda_R$; (b) Central frequency $\overline \omega d \sqrt{\rho^{PVDF}}/\sqrt{C_{3333}^{PVDF}}$ versus the tuning parameter $\lambda^*=\lambda-\lambda_R$.}
    \label{figure9}
\end{figure}

\noindent To sum up, it can be observed that the geometry of the metamaterial, i.e. the relative size of its elastic and shunting piezoelectric elements, has a significant impact on the operation of the device 
with or without triggering the shunting effect, that is for $\lambda  \approx \lambda_R$  and $\lambda$=0.\\
A further investigation has been performed by comparing the results obtained for the 
three geometries considered so far, in terms of 
both dimensionless band gap amplitude $A_s^+ d \sqrt{\rho^{PVDF}}/\sqrt{C_{3333}^{PVDF}}$ and central frequency $\overline \omega d \sqrt{\rho^{PVDF}}/\sqrt{C_{3333}^{PVDF}}$ versus the tuning parameter $\lambda^*=\lambda-\lambda_R$.
Attention is focused on the first band gap forming between the 1$^{st}$  and the 2$^{nd}$ optical branches of the Floquet-Bloch spectrum. \textcolor{black}{ In Fig. 8 the blue curve refers to the ratio $R/d$=0.3, the red curve to $R/d$=0.35, and the black curve refers to $R/d$=0.4.} It can be noted irrespective of the considered geometry, the band gap amplitude, Fig. \ref{figure7}(a), and the central frequency, Fig. \ref{figure7}(b), increase as $\lambda^*$ decreases, i.e. when $\lambda$ tends to $\lambda_R$. 
The maximum amplitude is guaranteed by the metamaterial with the smallest radii $r$ and $R$. Moreover, a peculiar trend is observed for $R/d$=0.4, since in this case it is possible to open a band gap and controlling its amplitude and central frequency by intervening on the tuning  parameter $\lambda$, starting from a dense spectrum for $\lambda$=0.
The analysis of additional geometries characterizing the metamaterial reveals that the use of ratios $r/R$=0.7 and $R/d$=0.4 ensures a wide range of behaviours as $\lambda$ changes. In Fig. \ref{figure8} the Floquet- Bloch spectrum referred to this geometry is shown for different values of $\lambda$. Specifically, Fig. \ref{figure8}(a) shows the curve related to $\lambda$=0. Note that in this case, where the shunting effects are not triggered, a wide total band gap is detected between the 1$^{st}$  and the 2$^{nd}$ optical branches. As $\lambda$ decreases further band gaps opens, as in Fig. \ref{figure8}(b) corresponding to $\lambda$=-1.0100, in which a new total band gap forms between the 15$^{th}$ and the 16$^{th}$ optical branches. Moving towards a lower value of $\lambda$=-1.0334, Fig. \ref{figure8}(c), besides the first band gap, three additional total band gaps open between the 2$^{nd}$- 3$^{rd}$, 5$^{th}$-6$^{th}$, 12$^{th}$-13$^{th}$ optical branches, respectively. \textcolor{black}{Finally, when $\lambda_R$ is approached as in Fig. \ref{figure8}(d) with $\lambda$=-1.0368, the first band gap remains almost unchanged, while the two additional  total band gaps at low frequencies, between the 2$^{nd}$- 3$^{rd}$, 4$^{th}$-5$^{th}$ optical branches, becomes wider}. In the considered frequency range the spectrum tends to become less dense as $\lambda$ decreases. \textcolor{black}{Moreover, in Fig. \ref{figure8a} the wave forms  of the Floquet-Bloch spectrum for the metamaterial with $r/R$=0.7 and $R/d$=0.4 for $\lambda  \approx \lambda_R$, related to relevant points highlighted in Figure \ref{figure8}(d), are reported. In particular, the deformed shape together with the contour plot related to the dimensionless quantity $\lVert$Re$\left( \boldsymbol{\mathord{\buildrel{\lower3pt\hbox{$\scriptscriptstyle\frown$}} \over u}} \right)\rVert_2/d$ are shown.} 
Also in this case we investigate the behaviour of the metamaterial in terms of 
both dimensionless band gap amplitude $A_s^+ d \sqrt{\rho^{PVDF}}/\sqrt{C_{3333}^{PVDF}}$, Fig. \ref{figure9}(a), and central frequency $\overline \omega d \sqrt{\rho^{PVDF}}/\sqrt{C_{3333}^{PVDF}}$,  Fig. \ref{figure9}(b), versus the tuning parameter $\lambda^*=\lambda-\lambda_R$.
.
\noindent The first two lower band gaps, forming between the 1$^{st}$-2$^{nd}$ and 2$^{nd}$-3$^{rd}$ optical branches, are analysed. In particular, the blue curves refer to the lowest frequency total band gap, while the red curves are referred to the other band gap.
The band gap amplitude, associated to the blue curve in Fig. \ref{figure9}(a), tends to increase as $\lambda^*$ decreases. As expected,  analogous behaviour is detected for the corresponding central frequency in Fig. \ref{figure9}(b). On the other hand, concerning the red curves, an interesting behaviour is observed related to the 
formation of a new total band gap from a dense spectrum

\subsection{\textcolor{black}{Filtering performances of the tunable metamaterial}}
\label{Section41}
\begin{figure}[H]
\centering
\includegraphics[width=1\textwidth]{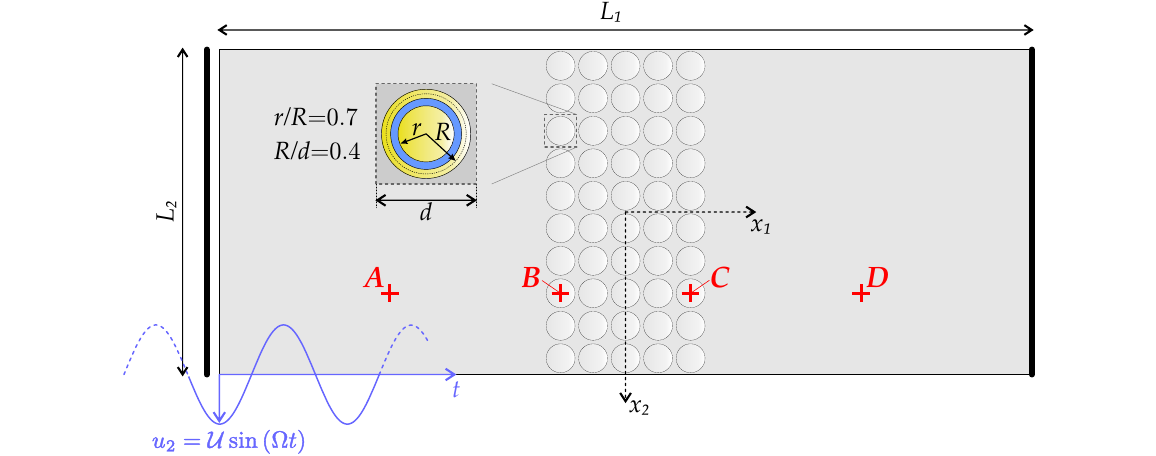}
\caption{\footnotesize Thin rectangular strip of homogeneous material with a central core made by a portion of the metamaterial, undergoing a mono-frequency time-harmonic in-plane displacement excitation.}
\label{figure410}
\end{figure}

\begin{figure}[h!]
\centering
\includegraphics[width=1\textwidth]{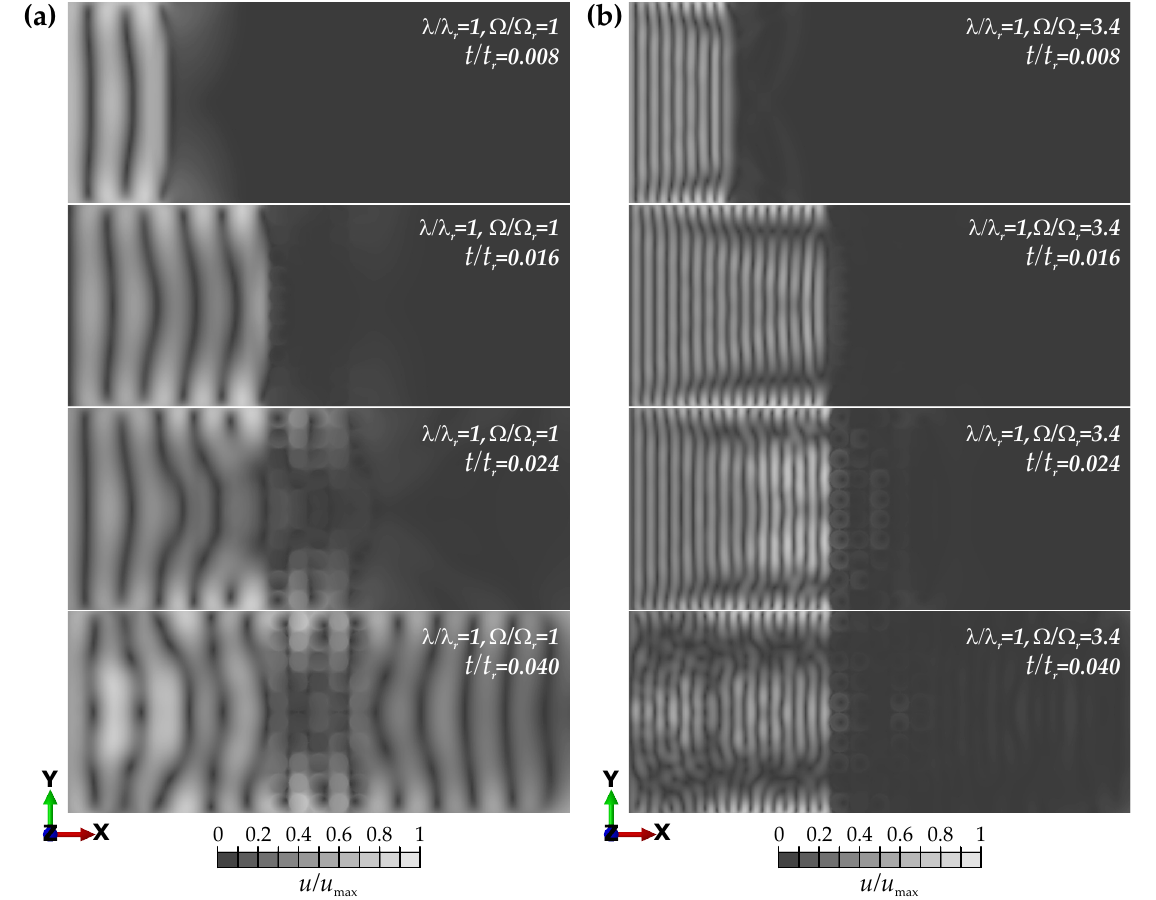}
\caption{ \footnotesize Contour plots of the dimensionless displacement magnitudes $u/u_{max}$ at different dimensionless time instants $t/t_r$. (a)  dimensionless excitation frequency $\Omega/\Omega_r$=1; (b)  dimensionless excitation frequency $\Omega/\Omega_r$=3.4. }
\label{figure411}
\end{figure}

In order to assess the filtering properties of the proposed tunable metamaterial we here present a numerical experiment developed in Abaqus.
\textcolor{black}{We consider the specimen depicted in Fig. \ref{figure410}, made up of a thin rectangular strip of homogeneous material in which a central stripe has been replaced by a cluster of ten by five periodic cells of the designed metamaterial. The specimen has dimensions $L_1$= 250 mm, $L_2$=100 mm and uniform thickness $w$, while the geometry of the periodic cell is the same adopted in Fig. \ref{figure8}, i.e. characterized by $r/R$=0.7 and  $R/d$=0.4, with the edge $d$=10 mm.} More specifically, the homogeneous material is made of epoxy resin (EPO-TEK$^\text{\textregistered}$ 301) such as the soft matrix of the metamaterial (see Sect. \ref{Section4}).
\noindent 
The left and right sides of the specimen are fully clamped, while the bottom and top sides are subject to free boundary conditions.
 As excitation source, an in-plane mono-frequency time-harmonic displacement excitation $u_2=\mathcal{U}$sin$(\Omega t)$
 has been imposed to the left side, being $\mathcal{U}$=0.01 mm the excitation amplitude and $\Omega$ its angular frequency.
 The geometry of the sample has been designed in SolidWorks and then imported in Abaqus. The specimen has been discretized using S4R elements. Mesh size and time step were carefully chosen accordingly to the Courant-Friedrichs-Lewy (CFL) condition and Blake's criteria \citep{CFL,de2013courant}. More specifically, the higher is the frequency, the smaller has to be the mesh size and the time step. \textcolor{black}{In particular, we performed a mesh-refinement study to estimate the optimal mesh-size so that the results to be mesh-independent.} The numerical simulation of the undamped dynamic response has been achieved by means of an dynamic/explicit  time-stepping scheme.\\
 Depending on the excitation frequency, qualitatively different behaviours 
 are observed.  More specifically, in the case shown in Fig. \ref{figure411}(a), i.e. when the dimensionless excitation frequency $\Omega/\Omega_r$, being $\Omega_r=(d\sqrt{\rho^{PVDF}/C_{3333}^{PVDF} })^{-1} $,  falls within the first low-frequency pass band highlighted in Fig. \ref{figure8}(d) in black/dashed line, \textcolor{black}{with the tuning parameter $\lambda  \approx \lambda_R$,} a marked propagation of acoustic waves through the metamaterial core is found.

\begin{figure}[H]
	\centering
	\includegraphics[width=1\textwidth]{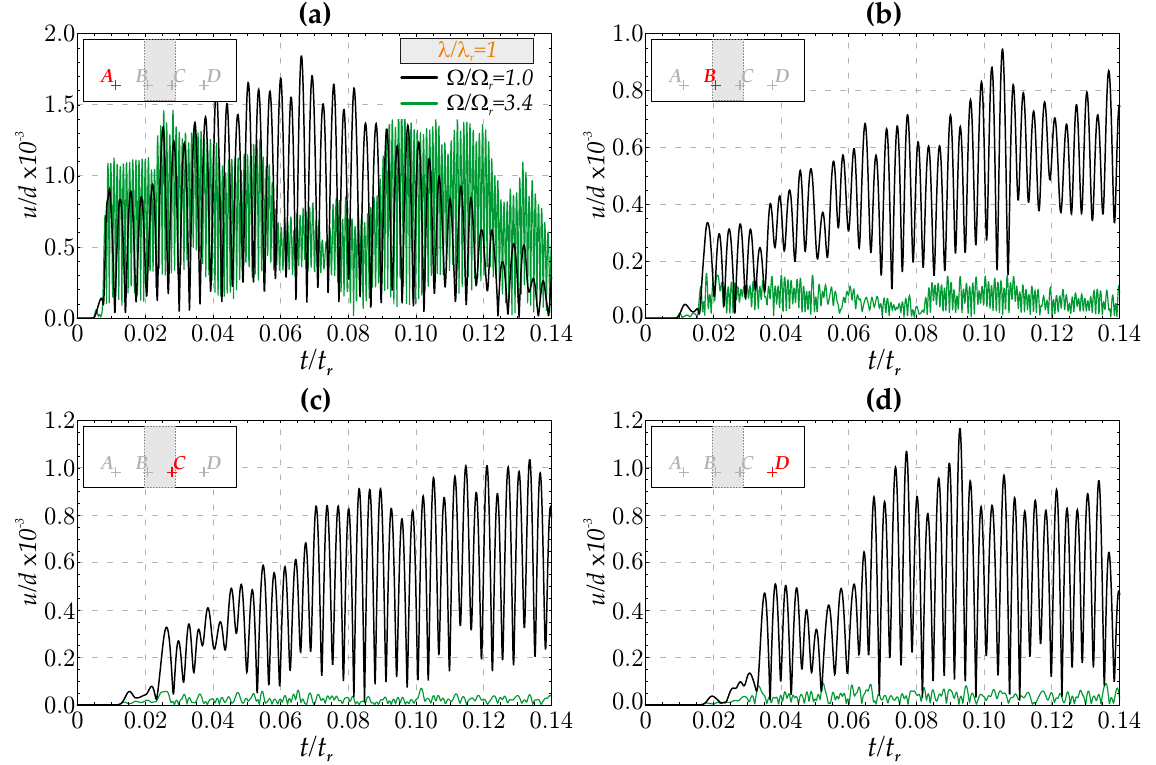}
	\caption{\footnotesize Time histories of the dimensionless displacement magnitudes $u/d$ versus the dimensionless time $t/t_r$ for two characteristic values of the dimensionless excitation frequency $\Omega/\Omega_r$ in four control points: (a) point $A$; (b) point $B$; (c) point $C$; (d) point $D$.}
	\label{figure420}
\end{figure}
\noindent Looking at the contour plots of the dimensionless displacement magnitudes $u/u_{max}$, defined as the ratio between $u=\|u_i \textbf{e}_i \|_2$, with  i=1,2, and the respective maximum value $u_{max}$, that are referred to different instants of time  $t/t_r$, with $t_r=d \sqrt{\rho^{PVDF}/C_{3333}^{PVDF} }$,
it emerges, indeed, that the wave-front propagates starting from the left side and at the dimensionless time about $t/t_r=0.016$ it reaches the microstructured core. After various diffraction, reflection and refraction phenomena, the advancing 
wave-front propagates through the metamaterial at dimensionless times  greater than $t/t_r=0.024$ without noticeable attenuation in the excitation amplitude. \\
The wave propagation is also confirmed by the time histories of the dimensionless displacement magnitudes $u/d$, black curves in Fig. \ref{figure420}, measured at equispaced control points. All the points are characterized by displacement magnitudes quantitatively comparable with each other, this means that the metamaterial does not
give rise to wave trapping effect, so that it does not act as acoustic filter.
On the other hand, in the case depicted  in Fig. \ref{figure411}(b), i.e. when the excitation frequency falls within the first low-frequency stop band, highlighted in Fig. \ref{figure8}(d) in green/dash-dot line, \textcolor{black}{with the tuning parameter $\lambda  \approx \lambda_R$,} 
the metamaterial core is definitely able to act as a filter. In this case, by observing the contour plots of the dimensionless displacement magnitudes $u/u_{max}$,  
it is evident that, as in the previous case, the wave-front propagates starting from the left side and at about the same dimensionless time $t/t_r=0.016$ it reaches the microstructured core, giving rise to diffraction, reflection and refraction phenomena.  
Conversely, the propagating acoustic wave is trapped in the metamaterial core, that works as a metafilter, as shown at dimensionless times  greater than $t/t_r=0.024$.
Also in this case, this behaviour is confirmed by the time histories of the dimensionless displacement magnitudes $u/d$, green curves in Fig. \ref{figure420}, measured at the same equispaced control points. By comparing the displacement magnitudes pertaining to points $A$ and $D$, it is noted that the latter values are almost negligible with respect to the former ones. Moreover, as expected 
the displacement magnitudes in $C$ is noticeably lower than that in $B$.\\
\textcolor{black}{A further investigation aimed at showing the filtering performances of the tunable metamaterial as the tuning parameter changes is also performed. 
In particular, in the case shown in Fig. \ref{figure411N}(a), i.e. 
for tuning parameter $\lambda $=0 and dimensionless excitation frequency $\Omega/\Omega_r=5$, falling within the first low-frequency pass band highlighted in Fig. \ref{figure8}(a) in magenta/dash-double dot line, a marked propagation of acoustic waves through the metamaterial core is found.
\begin{figure}[h!]
\centering
\includegraphics[width=1\textwidth]{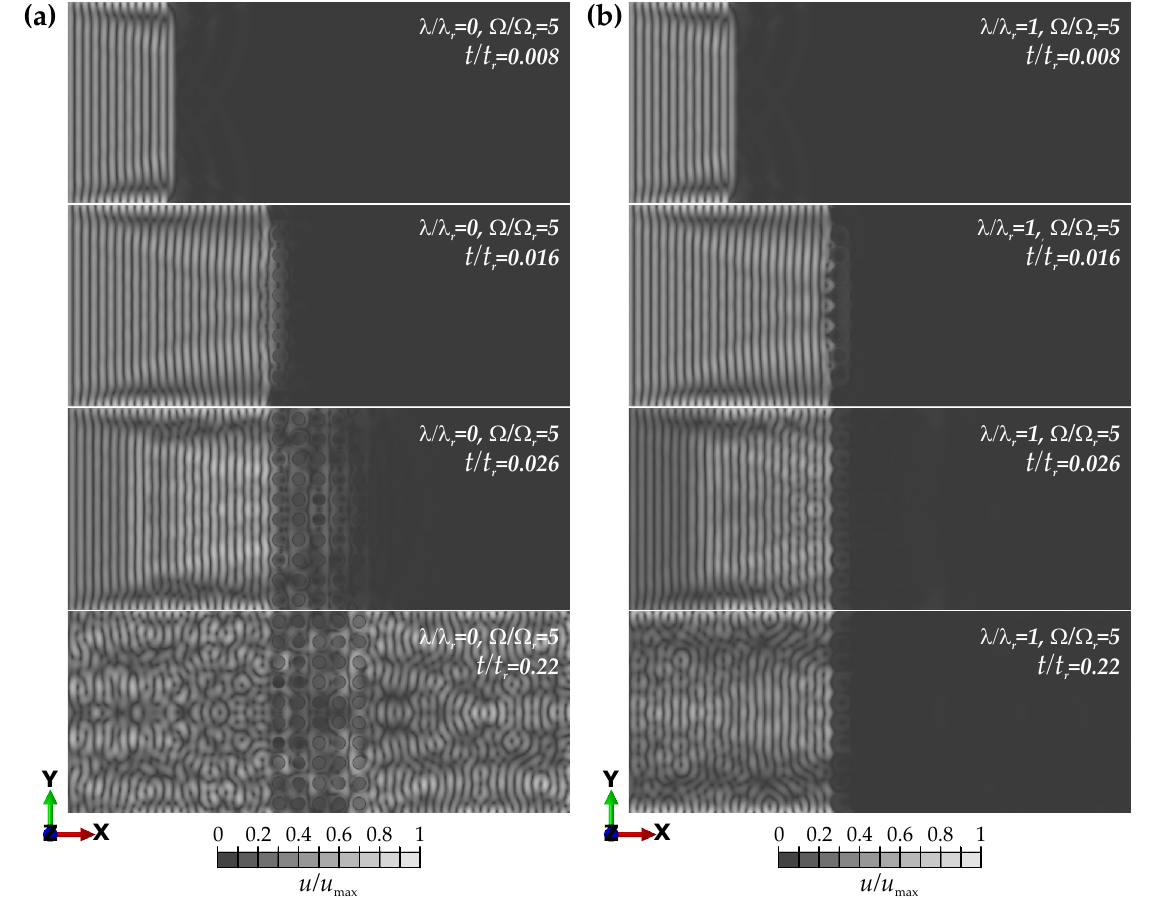}
\caption{ \textcolor{black}{\footnotesize Contour plots of the dimensionless displacement magnitudes $u/u_{max}$ at different dimensionless time instants $t/t_r$ for dimensionless excitation frequency $\Omega/\Omega_r$=5. (a) $\lambda$=0; (b) $\lambda  \approx \lambda_R$. }}
\label{figure411N}
\end{figure}
\begin{figure}[h!]
	\centering
	\includegraphics[width=1\textwidth]{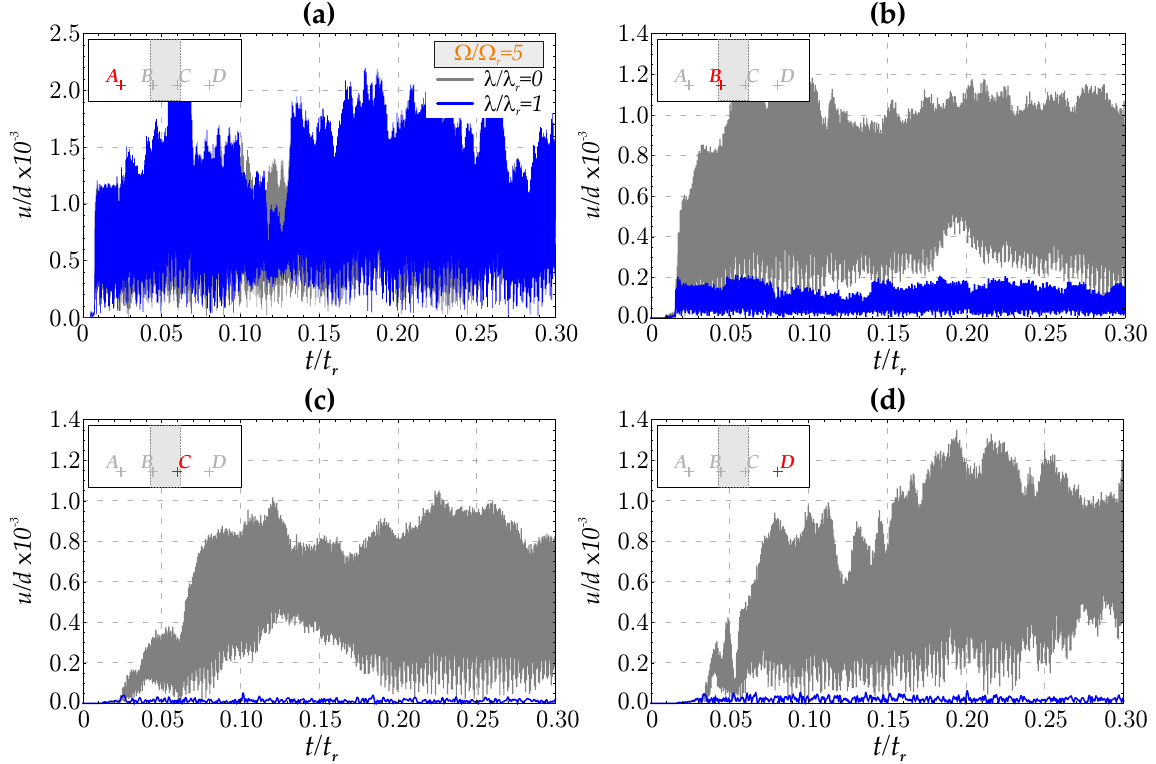}
	\caption{\textcolor{black}{\footnotesize Time histories of the dimensionless displacement magnitudes $u/d$ versus the dimensionless time $t/t_r$ for two characteristic values of the tuning parameter $\lambda$ and for dimensionless excitation frequency $\Omega/\Omega_r$=5  in four control points: (a) point $A$; (b) point $B$; (c) point $C$; (d) point $D$.}}
	\label{figure420N}
\end{figure}
Considering the contour plots of the dimensionless displacement magnitudes $u/u_{max}$, referred to different instants of time  $t/t_r$,
it appears that the wave-front propagates from the left side and at the dimensionless time about $t/t_r=0.016$ it reaches the microstructured core. After various diffraction, reflection and refraction phenomena, the advancing 
wave-front propagates through the metamaterial at dimensionless times  greater than   $t/t_r=0.026$ without noticeable attenuation in the excitation amplitude. 
Also in this case, the wave propagation is confirmed by the time histories of the dimensionless displacement magnitudes $u/d$, grey curves in Fig. \ref{figure420N}, measured at equispaced control points. All the points are characterized by displacement magnitudes quantitatively comparable with each other, i.e. the metamaterial does not act as acoustic filter.\\
\noindent Conversely, in the case depicted  in Fig. \ref{figure411N}(b), i.e. for tuning parameter $\lambda \approx \lambda_R$ and dimensionless excitation frequency $\Omega/\Omega_r=5$, falling within the second frequency stop band, highlighted in Fig. \ref{figure8}(d) in magenta/dash-double dot line, 
the metamaterial core is certainly able to act as a filter. Also in this case, by observing the contour plots of the dimensionless displacement magnitudes $u/u_{max}$,  
it is evident that the wave-front propagates from the left side and at about the same dimensionless time $t/t_r=0.016$ it reaches the microstructured core, giving rise to diffraction, reflection and refraction phenomena. 
On the other hand, the propagating acoustic wave is trapped in the metamaterial core, working as a metafilter, as shown at dimensionless times  greater than $t/t_r=0.026$.
This behaviour is again confirmed by the time histories of the dimensionless displacement magnitudes $u/d$, blue curves in Fig. \ref{figure420N}, measured at the same equispaced control points. By comparing the displacement magnitudes pertaining to points $A$ and $D$, it is noted that the latter values are almost negligible with respect to the former ones. Moreover, as expected 
the displacement magnitudes in $C$ is noticeably lower than that in $B$.}
\textcolor{black}{A couple of videos showing the filtering performances of the tunable metamaterial 
are available as supplementary material.
}

\section{Final remarks}\label{S5}
A  class of tunable acoustic metamaterials with periodic piezoelectric microstructure, shunted by an electrical circuit and characterized by adjustable frequency band structure is here proposed. The main goal is the design of  high-performance acoustic filters for the adaptive passive control of wave propagation 
in periodic materials. Attention is paid to three-phase microstructured materials made of  a phononic crystal coupled to local resonators. In particular,  the periodic phononic crystal is characterized by an in-plane regular repetition of rigid and heavy external rings embedded within a soft and light matrix. On the other hand, the local resonators are made of rigid and heavy internal disks connected to the external rings through inner rings made of a piezoelectric material shunted by an electrical circuit.
In the framework of a micromechanical approach, the dispersive waves propagation within the periodic medium has been analysed, by exploiting a generalization of the Floquet-Bloch theory, in order to determine the frequency band structure of such materials. The  governing equations of the in-plane Bloch-wave propagation have been introduced, complemented by their associated Floquet-Bloch boundary conditions. The constitutive equations of the orthotropic shunted piezoelectric material have been derived in the Laplace transformed space. In the case the piezoelectric phase is shunted by a generic dissipative electrical circuit, the components of the constitutive tensors are in general dependent on the complex frequency, contrary to what happens when a purely capacitive non dissipative electrical circuit is considered. 
In particular, the time domain governing equations of the in-plane free Bloch wave propagation turn out to be formally analogous to either those of a periodic  viscoelastic continuum in the case of a dissipative electrical circuit, or analogous to those of a periodic elastic continuum in the case of a non dissipative electrical circuit. It follows that in the former case a complex-value frequency band structure is observed, while in the latter case a real-value one.
Focus is on metamaterial shunted by electrical circuits characterized by equivalent purely capacitive admittance, in which the positive or negative-value capacitance can be externally modified.
 It follows that the changeable capacitance plays the role of a tuning parameter and is directly used to adaptively control the constitutive elastic properties of the piezoelectric shunting material, as well as, the resulting acoustic properties of the tunable metamaterial. In particular, attention is paid to tuning parameter ranges corresponding to positive values of the constitutive components of the elastic tensors. The influence of geometric design parameters on the Floquet-Bloch spectra of the acoustic metamaterials is investigated, as the tuning parameter changes. It is demonstrated that it is possible to open low-frequencies band gaps and controlling their amplitude and central frequencies by intervening on the tuning parameter, also starting from dense spectra. More specifically, by decreasing the tuning parameter up to about the  $\textit{resonant}$ value, a stiffening of  the piezoelectric material is observed, together with enlargements of low-frequency band gaps. Moreover, as expected the central frequencies of such band gaps tend to slightly increase. Finally, the effectiveness of the periodic tunable metamaterial
as acoustic filter has been tested through a a numerical experiment on a thin rectangular strip of homogeneous material in which a central strip has been replaced by a portion of the metamaterial. By imposing a mono-frequency time-harmonic displacement excitation to the left side of the specimen, it results that, as expected, when the excitation frequency falls into a pass band no filtering properties are exhibited. On the other hand, when the excitation frequency falls into a stop band, the tunable metamaterial behaves as a highly efficient acoustic filter.
The promising results obtained can be positively exploited towards optimal design of tunable acoustic filters, adapt to a changing performance requirement in real-time.\\
Interesting future developments may concern, on the one hand, the study of the acoustic behavior of tunable metamaterials connected to more sophisticated and even dissipative electrical circuits and, on the other, the design of tunable metamaterials with smart and/or imperfect microstructured interfaces \citep{zhou2012bulk,sevostianov2012connections,massabo2020local,monetto2019effects}.

\section*{Acknowledgements}
{The authors acknowledge the financial support from National Group of Mathematical Physics (GNFM-INdAM). D.M is supported by the European Commission under the FET Open (Boheme) grant No. 863179.
}

\bibliographystyle{apalike}
\bibliography{BIBLIO}

\textcolor{black}{\section*{Appendix A: Constitutive equations for the three-dimensional orthotropic piezoelectric material}
The coupled
constitutive relations for the three-dimensional orthotropic piezoelectric material with polarization along $\textbf{e}_3$
 in the stress-charge form read
\begin{align}
&\mathord{ \sigma}_{ij}=C_{ijhk} \, \mathord{ \varepsilon}_{hk}+C_{ij33} \, \mathord{ \varepsilon}_{33}+ e_{ij3} \, \frac{\partial \mathord{ \phi}}{\partial x_3},\nonumber \\
&\mathord{ \sigma}_{33}=C_{33hk} \, \mathord{ \varepsilon}_{hk}+C_{3333} \, \mathord{ \varepsilon}_{33}+ e_{333} \,  \frac{\partial \mathord{ \phi}}{\partial x_3},\nonumber\\
&\mathord{ \sigma}_{\alpha3}=2 \, C_{\alpha3\alpha3} \, \mathord{ \varepsilon}_{\alpha3}+ e_{\alpha3\alpha} \,  \frac{\partial \mathord{ \phi}}{\partial x_l},\nonumber \\
&\mathord{ D}_{i}=2 \, \widetilde{e}_{ij3} \, \mathord{ \varepsilon}_{j3}+ \beta_{ij} \,  \frac{\partial \mathord{ \phi}}{\partial x_j},\nonumber \\
&\mathord{ D}_{3}= \widetilde{e}_{3jh} \, \mathord{ \varepsilon}_{jh}+ \widetilde{e}_{333} \, \mathord{ \varepsilon}_{33}- \beta_{33} \,  \frac{\partial \mathord{ \phi}}{\partial x_3}, \qquad i,j,h,k,\alpha=1,2,\label{30}
\end{align}
where from now on no summation on index $\alpha$ is applied, and being $\sigma_{pq}$ the components of the stress tensor, $D_{p}$ the components of the 
 electric displacement field, $\varepsilon_{rs} = (\partial u_r/ \partial x_s+ \partial u_s /\partial x_r) / 2$ is the components of the strain tensor, with $u_p$ the  displacement field components, $\phi$ the  electric potential field, 
 $ C_{pqrs}$ the components of the fourth order elasticity tensor, 
$\beta_{pr}$
the components of the second order dielectric
permittivity tensor, ${e}_{qsp}$ the components of  the third order piezoelectric stress-charge coupling tensor and its transpose $\widetilde{e}_{pqs}={e}_{qsp}$. \\
For the sake of convenience we apply the bilateral Laplace transform to the governing equations (\ref{30}). By recalling the definition of the bilateral Laplace transform for a generic  function $g(\textbf{x},t)$, i.e.
\begin{align} \label{23}
\mathcal{L} \left[ {g}(\textbf{x},t) \right]= \int_{-\infty}^{+\infty} {g}(\textbf{x},t)e^{-s t}d{t}=\mathord{\buildrel{\lower3pt\hbox{$\scriptscriptstyle\frown$}} \over g} (\textbf{x},s),
\end{align}
being $s \in \mathbb{C}$ the Laplace variable playing the role of the complex angular frequency,
 the transformed coupled 
constitutive relations in the stress-charge form,  read 
\begin{align}
&\mathord{\buildrel{\lower3pt\hbox{$\scriptscriptstyle\frown$}} \over \sigma}_{ij}=C_{ijhk} \, \mathord{\buildrel{\lower3pt\hbox{$\scriptscriptstyle\frown$}} \over \varepsilon}_{hk}+C_{ij33} \, \mathord{\buildrel{\lower3pt\hbox{$\scriptscriptstyle\frown$}} \over \varepsilon}_{33}+ e_{ij3} \, \frac{\partial \mathord{\buildrel{\lower3pt\hbox{$\scriptscriptstyle\frown$}} \over \phi}}{\partial x_3},\nonumber \\
&\mathord{\buildrel{\lower3pt\hbox{$\scriptscriptstyle\frown$}} \over \sigma}_{33}=C_{33hk} \, \mathord{\buildrel{\lower3pt\hbox{$\scriptscriptstyle\frown$}} \over \varepsilon}_{hk}+C_{3333} \, \mathord{\buildrel{\lower3pt\hbox{$\scriptscriptstyle\frown$}} \over \varepsilon}_{33}+ e_{333} \,  \frac{\partial \mathord{\buildrel{\lower3pt\hbox{$\scriptscriptstyle\frown$}} \over \phi}}{\partial x_3},\nonumber\\
&\mathord{\buildrel{\lower3pt\hbox{$\scriptscriptstyle\frown$}} \over \sigma}_{\alpha3}=2 \, C_{\alpha3\alpha3} \, \mathord{\buildrel{\lower3pt\hbox{$\scriptscriptstyle\frown$}} \over \varepsilon}_{\alpha3}+ e_{\alpha3\alpha} \,  \frac{\partial \mathord{\buildrel{\lower3pt\hbox{$\scriptscriptstyle\frown$}} \over \phi}}{\partial x_\alpha},\nonumber \\
&\mathord{\buildrel{\lower3pt\hbox{$\scriptscriptstyle\frown$}} \over D}_{i}=2 \, \widetilde{e}_{ij3} \, \mathord{\buildrel{\lower3pt\hbox{$\scriptscriptstyle\frown$}} \over \varepsilon}_{j3}+ \beta_{ij} \,  \frac{\partial \mathord{\buildrel{\lower3pt\hbox{$\scriptscriptstyle\frown$}} \over \phi}}{\partial x_j},\nonumber \\
&\mathord{\buildrel{\lower3pt\hbox{$\scriptscriptstyle\frown$}} \over D}_{3}= \widetilde{e}_{3jh} \, \mathord{\buildrel{\lower3pt\hbox{$\scriptscriptstyle\frown$}} \over \varepsilon}_{jh}+ \widetilde{e}_{333} \, \mathord{\buildrel{\lower3pt\hbox{$\scriptscriptstyle\frown$}} \over \varepsilon}_{33}- \beta_{33} \,  \frac{\partial \mathord{\buildrel{\lower3pt\hbox{$\scriptscriptstyle\frown$}} \over \phi}}{\partial x_3}, \qquad i,j,h,k,\alpha=1,2.\label{31}
\end{align}
The transformed electric potential difference $\mathord{\buildrel{\lower3pt\hbox{$\scriptscriptstyle\frown$}} \over V}_{3}$, imposed between the two opposite external surfaces of the piezoelectric ring, orthogonal to $\textbf{e}_3$  as shown in Fig. \ref{figurePrima}(b), and the transformed current $\mathord{\buildrel{\lower3pt\hbox{$\scriptscriptstyle\frown$}} \over I}_{3}$ along the $\textbf{e}_3$ direction are defined in the Laplace domain, in agreement with 
\cite{HAGOOD1991}, as
\begin{align}
&\mathord{\buildrel{\lower3pt\hbox{$\scriptscriptstyle\frown$}} \over V}_{3}= -L^{(P)}  \frac{\partial \mathord{\buildrel{\lower3pt\hbox{$\scriptscriptstyle\frown$}} \over \phi}}{\partial x_3} ,\nonumber \\
&\mathord{\buildrel{\lower3pt\hbox{$\scriptscriptstyle\frown$}} \over I}_{3}=s \, A^{(P)} \mathord{\buildrel{\lower3pt\hbox{$\scriptscriptstyle\frown$}} \over D}_3 
   ,\label{32}
\end{align}
where $ A^{(P)}= \pi ((R-h/2)^2-r^2)$ is the area of the piezoelectric annulus, and $L^{(P)}=w$.
By replacing the equations (\ref{32}) into the constitutive relations (\ref{31}), 
the transformed
coupled constitutive relations (\ref{33}), expressed in terms of $\mathord{\buildrel{\lower3pt\hbox{$\scriptscriptstyle\frown$}} \over V}_{3}$, $\mathord{\buildrel{\lower3pt\hbox{$\scriptscriptstyle\frown$}} \over I}_{3}$ are obtained.}

\section*{Appendix B: Equivalent properties of the electrical circuit shunting the piezoelectric phase}

Considering a generic external electrical circuit,
the constitutive relations  in terms of the  electric potential difference $\mathord{\buildrel{\lower3pt\hbox{$\scriptscriptstyle\frown$}} \over V}_{3}$ and of the  current $\mathord{\buildrel{\lower3pt\hbox{$\scriptscriptstyle\frown$}} \over I}_{3}$ in the transformed Laplace space result
\begin{align}
&\mathord{\buildrel{\lower3pt\hbox{$\scriptscriptstyle\frown$}} \over I}_{3}=   \, Y^{(SU)}_{33}(s)  \, \mathord{\buildrel{\lower3pt\hbox{$\scriptscriptstyle\frown$}} \over V}_{3} .\label{34a}
\end{align}
where $Y^{(SU)}_{33}(s)$ is the equivalent admittance of the electrical circuit. By applying the inverse bilateral Laplace transform 
to (\ref{34a}), the following integral constitutive equation in the
time domain is obtained 
\begin{align}
&I_{3}= \mathcal{L}^{-1}  \left[  \,\frac{ Y^{(SU)}_{33}(s)}{s} \right]  \,* \frac{\partial V_{3}}{ \partial t}, \label{34b}
\end{align}
being $*$ is the convolution product. Note that equation  
(\ref{34b}) characterizes a generic non quasi-static 
electrical circuit (also referred as \textit{time-modulated})  or an electric circuit with non purely capacitive equivalent admittance  \citep{IntegroDiff}. \\
\textcolor{black}{On the other hand, in the case of a purely capacitive equivalent admittance, i.e. when the equivalent admittance linearly depends on the Laplace variable $s$, the constitutive equation in the time domain takes the well-known form
\begin{align}
&I_{3}= \mathcal{C} \delta(t) * \frac{\partial V_{3}}{ \partial t} = \mathcal{C}\frac{\partial V_{3}}{ \partial t} ,\label{34}
\end{align}
where $\delta (t)$ is the Dirac delta function. }

\end{document}